\documentclass[trackchanges, twocolumn, twocolappendix]{aastex701}
\usepackage{amsmath, physics}

\shorttitle{Constraint on the Optical Depth from the Reionization History}
\shortauthors{Kageura et al.}

\newcommand{\xHI}{x_\mathrm{H\,\textsc{i}}}
\newcommand{\revise}[1]{\textcolor{black}{#1}}

\begin{document}

\title{
A New Constraint on the Optical Depth from the Reionization History\\
Independent of CMB Large-Scale E-Mode Polarization


%
}

\author[0009-0004-0381-7216]{Yuta Kageura}
\affiliation{Institute for Cosmic Ray Research, The University of Tokyo, 5-1-5 Kashiwanoha, Kashiwa, Chiba 277-8582, Japan}
\affiliation{Department of Physics, Graduate School of Science, The University of Tokyo, 7-3-1 Hongo, Bunkyo, Tokyo 113-0033, Japan} 
\email{kageura@icrr.u-tokyo.ac.jp}

\author[0000-0002-1049-6658]{Masami Ouchi}
\affiliation{National Astronomical Observatory of Japan, 2-21-1 Osawa, Mitaka, Tokyo 181-8588, Japan}
\affiliation{Institute for Cosmic Ray Research, The University of Tokyo, 5-1-5 Kashiwanoha, Kashiwa, Chiba 277-8582, Japan}
\affiliation{Astronomical Science Program, Graduate Institute for Advanced Studies, SOKENDAI, 2-21-1 Osawa, Mitaka, Tokyo 181-8588, Japan}
\affiliation{Kavli Institute for the Physics and Mathematics of the Universe (Kavli IPMU, WPI), The University of Tokyo, 5-1-5 Kashiwanoha,
Kashiwa, Chiba, 277-8583, Japan}
\email{ouchims@icrr.u-tokyo.ac.jp}

\author[0000-0003-3542-6637]{Fumihiro Naokawa}
\affiliation{Department of Physics, Graduate School of Science, The University of Tokyo, 7-3-1 Hongo, Bunkyo, Tokyo 113-0033, Japan}
\affiliation{Research Center for the Early Universe, The University of Tokyo, Bunkyo-ku, Tokyo 113-0033, Japan}
\affiliation{Department of Physics, Tohoku University, Sendai, Miyagi 980-8578, Japan}
\affiliation{Departamento de F\'isica de la Tierra y Astrof\'isica, Universidad Complutense de Madrid, 28040 Madrid, Spain}
\email{fumihiro.naokawa@resceu.s.u-tokyo.ac.jp}

\author[0009-0008-0167-5129]{Hiroya Umeda}
\affiliation{Institute for Cosmic Ray Research, The University of Tokyo, 5-1-5 Kashiwanoha, Kashiwa, Chiba 277-8582, Japan}
\affiliation{Department of Physics, Graduate School of Science, The University of Tokyo, 7-3-1 Hongo, Bunkyo, Tokyo 113-0033, Japan}
\email{ume@icrr.u-tokyo.ac.jp}

\author[]{Akinori Matsumoto}
\affiliation{Institute for Cosmic Ray Research, The University of Tokyo, 5-1-5 Kashiwanoha, Kashiwa, Chiba 277-8582, Japan}
\affiliation{Department of Physics, Graduate School of Science, The University of Tokyo, 7-3-1 Hongo, Bunkyo, Tokyo 113-0033, Japan}
\email{matsumoto-akinori489@g.ecc.u-tokyo.ac.jp}

\author[0000-0002-6047-430X]{Yuichi Harikane}
\affiliation{Institute for Cosmic Ray Research, The University of Tokyo, 5-1-5 Kashiwanoha, Kashiwa, Chiba 277-8582, Japan}
\email{hari@icrr.u-tokyo.ac.jp}

\author[0009-0000-1999-5472]{Minami Nakane}
\affiliation{Institute for Cosmic Ray Research, The University of Tokyo, 5-1-5 Kashiwanoha, Kashiwa, Chiba 277-8582, Japan}
\affiliation{Department of Physics, Graduate School of Science, The University of Tokyo, 7-3-1 Hongo, Bunkyo, Tokyo 113-0033, Japan}
\email{nakanem@icrr.u-tokyo.ac.jp}

\author[0000-0002-8408-4816]{Tran Thi Thai}
\affiliation{National Astronomical Observatory of Japan, 2-21-1 Osawa, Mitaka, Tokyo 181-8588, Japan}
\email{thithai.tran@nao.ac.jp}

\correspondingauthor{Yuta Kageura}
\email{kageura@icrr.u-tokyo.ac.jp}

\begin{abstract}
Recent studies report a mild discrepancy between baryon acoustic oscillation (BAO) and cosmic microwave background (CMB) measurements within the $\Lambda$CDM framework. This discrepancy could be explained if the optical depth $\tau$ inferred from the CMB large-scale E-mode polarization is underestimated, which may be biased by foreground-subtraction or instrumental systematics. In this work, we present a determination of $\tau$ independent of the large-scale E-mode polarization, using the latest measurements of the redshift evolution of the neutral hydrogen fraction $\xHI(z)$, which is constrained by Lyman-$\alpha$ forest and damping-wing absorption measurements at $z\sim5$--$14$, based on ground-based optical and JWST observations. 
Combining $\xHI(z)$ with the Planck CMB power spectra excluding the large-scale E-mode polarization, \revise{we obtain $\tau=0.0552^{+0.0019}_{-0.0026}(\mathrm{stat.})^{+0.0075}_{-0.0049}(\mathrm{sys.})$, where the systematic uncertainty accounts for possible absorption-modeling effects in the inference of $\xHI(z)$.
This constraint is consistent with previous CMB results including the large-scale E-mode polarization.
With this measurement,} we resolve the degeneracy in the $\tau$--$\Omega_{\rm m}$ plane and find a $2.4\sigma$ tension with the DESI DR2 BAO results, thereby confirming the claimed mild discrepancy suggestive of physics beyond $\Lambda$CDM.
Finally, we derive an upper limit on the sum of neutrino masses, $\Sigma m_\nu < 0.0550\,(0.0717)\,{\rm eV}$ at the 95\%\,(99\%) confidence level.
This limit favors the normal mass ordering and, when combined with the lower limits from neutrino oscillation experiments, yields a further constraint, $\Sigma m_\nu = 0.0594_{-0.0007}^{+0.0113}\,{\rm eV}$.
However, the cosmological upper limit and the oscillation-based lower limit show a mild $2.2\sigma$ tension, providing an independent indication of possible physics beyond $\Lambda$CDM.

\end{abstract}



\section{Introduction}
The Cosmic Microwave Background (CMB) is one of the most powerful probes for determining cosmological parameters.
In the base-$\Lambda$CDM model, the temperature and polarization power spectra, as well as CMB lensing measurements, are described by five cosmological parameters and the CMB optical depth to reionization, $\tau$ \citep{planck18_6}.
The parameter $\tau$ quantifies the Thomson scattering of CMB photons by free electrons after cosmic reionization, with earlier (later) reionization history corresponding to a larger (smaller) value of $\tau$.
A precise determination of $\tau$ is crucial because $\tau$ and other cosmological parameters both affect the amplitude of the power spectra, leading to parameter degeneracies.\par
Large-scale CMB E-mode polarization anisotropies are highly sensitive to $\tau$ \citep{planck_int_47}.
However, measuring this signal is challenging because the expected CMB polarization signal is more than two orders of magnitude weaker than the TT (temperature) power spectrum.
Both instrumental systematic effects and foreground residuals must be suppressed to very low levels to accurately extract the true CMB signal \citep{planck18_1}.
These observational challenges are evident in the history of $\tau$ estimates.\par
The Wilkinson Microwave Anisotropy Probe (WMAP) mission was the first to constrain $\tau$ based on the TE (temperature-polarization cross) power spectrum, obtaining a large value of $\tau=0.166_{-0.071}^{+0.076}$ in its 1-yr results \citep{wmap03}.
Subsequent analyses reduced this value to $\tau=0.089\pm0.014$ in the final 9-yr results, incorporating the TE and EE (polarization) power spectra \citep{wmap12}.\par
The Planck mission further revised $\tau$ downward to $\tau=0.079\pm0.017$ in its 2015 results, which included large-scale TE and EE measurements, or to $\tau=0.063\pm0.014$ after incorporating CMB lensing \citep{planck15_13}.
In the Planck 2018 results, $\tau$ was determined to be $\tau=0.0544\pm0.0073$ using the small-scale TT, TE, EE power spectra, large-scale TT and EE power spectra, and CMB lensing measurements \citep{planck18_6}.\par
The large-scale polarization measurements from Planck may be affected by instrumental systematics \citep{planck15_46}, and efforts to mitigate these systematics are ongoing.
In the Planck 2018 analysis, the \texttt{SimAll} likelihood was used for the low-multipole (i.e., large-scale, $l<30$) polarization power spectrum \citep{planck18_6}.
To address known instrumental effects, the \texttt{SRoll2} likelihood \citep{delouis19} was developed.
Based on Planck PR4 data, an alternative likelihood, \texttt{LoLLiPoP}, was constructed \citep{tristram21, tristram22}.
These likelihoods yield slightly different $\tau$ estimates \revise{\citep{pagano20, debelsunce21, tristram24}}. Thus, measuring the low-multipole polarization from Planck is still challenging.
We also note that current ground-based CMB results, including those from ACT \citep{act25} and SPT \citep{spt25}, can be affected by such systematic effects in Planck, since they rely on Planck results for the large-scale.\par
Moreover, several recent studies have proposed that $\tau$ may take higher values than those inferred from Planck \citep{allali2025, jhaveri25, liu25, sailer25}.
Since $\tau$ and $\Omega_m$ are negatively correlated, a higher value, $\tau\sim0.09$, can alleviate the mild tension in $\Omega_m$ between the baryon acoustic oscillation (BAO) measurements from the Dark Energy Spectroscopic Instrument (DESI) DR2 and Planck within the $\Lambda$CDM model \citep{jhaveri25, sailer25}.
In addition, several studies extending beyond the standard scenario have shown that the low-multipole Planck results can be consistent with $\tau\sim0.09$ \citep{naokawa24, namikawa25, tan25}.\par
Taking these observational challenges into account, we are strongly motivated to measure $\tau$ independently of the CMB large-scale E-mode polarization.
In this paper, we focus on the redshift evolution of the neutral hydrogen fraction, $\xHI(z)$, derived from quasar (QSO) and galaxy observations at $z\gtrsim5$, as an alternative probe to constrain $\tau$ \revise{\citep{hazra20, paoletti21, paoletti25, elbers25, garciagallego25, sims25}}.
A higher $\xHI$ leads to stronger Ly$\alpha$ absorption in these spectra, allowing $\xHI(z)$ to be constrained from the observations.\par
At $z\sim5-6$, Gunn-Peterson absorption \citep{gunn65} in the Lyman-series (e.g., Ly$\alpha$) forest of QSO spectra provides constraints on $\xHI$.
For example, dark gaps and the fraction of dark pixels with no detectable transmission in the Lyman-series forests are used to constrain $\xHI(z)$ \citep{mcgreer15, zhu22, jin23, davies25}.
Although Gunn-Peterson absorption cannot probe $z\gtrsim6$ owing to saturation, much weaker Lyman-$\alpha$ damping-wing absorption due to natural broadening (i.e., the uncertainty principle in energy and time) can constrain higher neutral fractions, $\xHI\sim0.1-1$.
At $z\sim6-7$, Lyman-$\alpha$ damping-wing absorption in QSO and gamma-ray burst (GRB) spectra is used to measure $\xHI(z)$ \citep{totani06, totani14, wang20, greig22, greig24, durovcikova24, fausey25}.
The damping-wing absorption of Ly$\alpha$ emission from galaxies is also used to investigate $\xHI(z)$ through the evolution of the Ly$\alpha$ luminosity function (LF) \citep{wold22, umeda25a}, Ly$\alpha$ emitter (LAE) clustering \citep{sobacchi15, ouchi18, umeda25a}, and the Ly$\alpha$ equivalent width (EW) distribution \citep{mason18, mason19b, jung20, whitler20, bolan22}. 
The James Webb Space Telescope (JWST) has further enabled us to constrain reionization history at $z\gtrsim8$, through the detection of damping-wing absorption in Lyman-break galaxy (LBG) spectra and the redshift evolution of the Ly$\alpha$ EW distribution \citep{nakane24, tang24, jones25, kageura25, mason25, napolitano25, umeda25b}.\par
We need to combine $\xHI(z)$ with other cosmological probes to derive $\tau$ and some other cosmological parameters. 
\revise{In practice, recent studies using pre-JWST data have constrained the optical depth from the reionization history, finding $\tau\sim0.05$ \citep{hazra20, krishak21, paoletti21, paoletti25, sims25}.}
However, because they combined CMB measurements that include the large-scale E-mode polarization, these results are not independent of potential issues in the low-multipole CMB polarization.\par
\citet{elbers25} obtained $\tau=0.0492_{-0.0030}^{+0.0014}$ independently of the CMB by combining BAO measurements, Big Bang nucleosynthesis (BBN), and reionization history.
This result is consistent with the Planck results and smaller than $\tau\sim0.09$, which has been proposed as a possible resolution of the CMB-BAO tension.
However, because BAO measurements are not dependent on $\tau$, combining reionization history with BAO does not improve the constraint on $\Omega_m$.
Therefore, the combination of reionization history and CMB provides a more powerful probe to test whether the CMB-BAO tension in $\Omega_m$ exists, by solving the degeneracy between $\tau$ and $\Omega_m$.
Measurements that exclude BAO are also important to avoid the impact of potential systematic errors in BAO measurements.\par
In this study, therefore, we present constraints on $\tau$ by combining reionization history with CMB measurements that exclude the large-scale E-mode polarization data. 
Through this combination, we test whether the CMB-BAO tension persists without large-scale E-mode measurements, which may be biased by instrumental or foreground-subtraction systematics. We incorporate the latest constraints on reionization history derived from the redshift evolution of Ly$\alpha$ emission observed with JWST and Subaru, in contrast to previous studies, while carefully excluding duplicate measurements.\par
This combination can simultaneously tighten constraints on other cosmological parameters that are degenerate with $\tau$ in the CMB alone. In addition, we perform further analyses combining CMB, BAO, and reionization history data to study models beyond $\Lambda$CDM.
We consider dynamical dark energy, which has been proposed as a possible resolution of the CMB-BAO tension, and the sum of neutrino masses, $\Sigma m_\nu$, which is degenerate with $\tau$. We also investigate potential systematic errors in measurements of the reionization history, which can bias cosmological parameters inferred from this combination.\par
This paper is organized as follows.
Section \ref{sec:data} describes the data used to constrain cosmological parameters.
Section \ref{sec:reconstract} presents the reconstruction of the reionization history based on Lyman-$\alpha$ forest data and QSO/galaxy Lyman-$\alpha$ damping-wing absorption measurements.
In Section \ref{sec:lcdm}, we describe constraints on the base-$\Lambda$CDM model.
Section \ref{sec:dde} explores constraints on the dynamical dark energy model.
Finally, we present constraints on the sum of neutrino masses in Section \ref{sec:mnu}.
Section \ref{sec:summary} summarizes our findings.

\section{Data}\label{sec:data}
In this work, we primarily use constraints on cosmic reionization history derived from galaxy and quasar observations at $z\gtrsim5$ (Section \ref{sec:crh}) and CMB measurements (Section \ref{sec:cmb}).
Data from other cosmological observations and high-energy experiments (Sections \ref{sec:bao} and \ref{sec:neutrino_data}) are also used to compare with our results for the base-$\Lambda$CDM model and to jointly constrain extensions to the base-$\Lambda$CDM model.

\subsection{Constraints on Cosmic Reionization History}\label{sec:crh}
We use constraints on $\xHI$ at various redshifts derived from galaxy and quasar observations at $z\gtrsim5$ reported in the literature.
In addition to Ly$\alpha$ forest and GRB/QSO/LBG damping-wing absorption measurements, we incorporate the latest constraints on reionization history derived from the redshift evolution of Ly$\alpha$ emission observed with JWST and Subaru, which were not used in previous studies \citep{elbers25, paoletti25, sims25}.
We note that many $\xHI$ measurements are based on the same observational data.
For example, the LBG spectrum used in \citet{curtislake23} is included in the sample of \citet{umeda25b}.
Although these duplicated measurements were used in some previous studies, we avoid duplication and use the following latest results from the literature:
\begin{itemize}
    \item Ly$\alpha$ damping-wing absorption of GRBs at $z=5.9-6.3$ \citep{totani06, totani14, fausey25}
    \item Ly$\alpha$ damping-wing absorption of QSOs at $z=6.0-7.5$ \citep{greig22, durovcikova24}
    \item Ly$\alpha$ damping-wing absorption of LBGs at $z=5-13$ \citep{umeda25b}
    \item Clustering properties of Ly$\alpha$ emitters at $z=5-7$ \citep{umeda25a}
    \item Evolution of the Ly$\alpha$ luminosity function at $z=5-7$ \citep{wold22, umeda25a}
    \item Evolution of the Ly$\alpha$ EW distribution at $z=5-14$ \citep{mason19a, jung20, whitler20, bolan22, kageura25}
    \item Dark gaps in the Ly$\beta$ forest at $z=5.5-6.0$ \citep{zhu22}
    \item Fraction of dark pixels in the Lyman-series forests at $z=4.9-6.2$ \citep{davies25}
\end{itemize}
\revise{We confirm that the selection of datasets does not significantly affect the result (see Appendix \ref{sec:TM} for details).}
Although the reionization history is well constrained at $z=5-8$, constraints are weaker at higher redshifts, especially at $z>10$.
To ensure that the reionization history remains a monotonically increasing function, we assume $\xHI>0.999$ ($1\sigma$ lower limit) at $z=15.0$, following \citet{elbers25}.
We note that the neutral fraction at $z=15$ is close to unity in most reionization models \citep{ishigaki18, finkelstein19, mason19b, naidu20, munoz24, asthana25, kageura25, qin25}.

\subsection{Cosmic Microwave Background}\label{sec:cmb}
\begin{deluxetable}{c|ccc}
    \tablecolumns{4}
    \tablewidth{\linewidth}
    \tabletypesize{\scriptsize}
    \tablecaption{CMB likelihoods used in this work.
    \label{tab:cmb}}
    \tablehead{
    Likelihood & Power spectrum & Name in \texttt{Cobaya} & Reference}
    \startdata
    \texttt{CamSpec} & Planck small-scale ($l\geq30$) TT+TE+EE & planck\_NPIPE\_highl\_CamSpec.TTTEEE & (1, 2)\\
    \texttt{Commander} & Planck large-scale ($l<30$) TT & planck\_2018\_lowl.TT\_clik & (3)\\
    Lensing & ACT+Planck CMB lensing & act\_dr6\_lenslike.ACTDR6LensLike & (4, 5, 6)
    \enddata
    \tablecomments{References. (1)\citet{efstathiou21}, (2)\citet{rosenberg22}, (3)\citet{planck18_5}, (4)\citet{carron22}, (5)\citet{madhavacheril24}, (6)\citet{qu24}}
\end{deluxetable}
We use publicly available CMB datasets (Table \ref{tab:cmb}).
Large-scale E-mode polarization data are excluded from our analysis because instrumental systematic effects and foreground residuals could significantly impact these measurements \citep{planck18_1, sailer25}.
For the small-scale ($l\ge30$) temperature (TT), polarization (EE), and cross (TE) power spectra, we use the \texttt{CamSpec} likelihood based on the Planck PR4 (NPIPE) maps \citep{efstathiou21, rosenberg22}.
For the large-scale ($l<30$) temperature power spectrum, we use the \texttt{Commander} likelihood from the Planck PR3 release \citep{planck18_5}\footnote{The PR3 and PR4 likelihoods are publicly available as internal likelihoods in \texttt{Cobaya} (\url{https://github.com/CobayaSampler/cobaya})}.
For CMB lensing, we use the v1.2 lensing likelihood from the ACT Collaboration, which combines the ACT DR6 and Planck PR4 results \citep{carron22, madhavacheril24, qu24}\footnote{\url{https://github.com/ACTCollaboration/act_dr6_lenslike}}.
Hereafter, we denote this dataset as CMB (w/o lowE).
In contrast, we denote the dataset including the large-scale E-mode power spectrum as CMB (w/ lowE).

\subsection{BAO}\label{sec:bao}
We use the DESI DR2 BAO likelihood \citep{desidr2_2, desidr2_1}\footnote{\url{https://data.desi.lbl.gov/public/papers/y3/bao-cosmo-params/README.html}}.
The combination of all subsets of the DESI BAO data is included in our analysis.

\subsection{Neutrino Oscillations}\label{sec:neutrino_data}
Results from neutrino oscillation experiments are used to jointly constrain the sum of neutrino masses $\Sigma m_\nu$.
\citet{esteban24} derived the neutrino mass-squared differences, $\Delta m_{21}^2$ and $\Delta m_{31}^2$, using various experimental data, including SK1-4, T2K, and NoVA, through NuFIT 6.0\footnote{\url{http://www.nu-fit.org}}.
We use the two-dimensional $\Delta\chi^2$ values for $\Delta m_{21}^2$ and $\Delta m_{31}^2$ to jointly constrain $\Sigma m_\nu$ in combination with reionization history, CMB, and BAO data.

\section{Reconstructing Reionization History}\label{sec:reconstract}
\begin{figure}
    \centering
    \includegraphics[width=\linewidth]{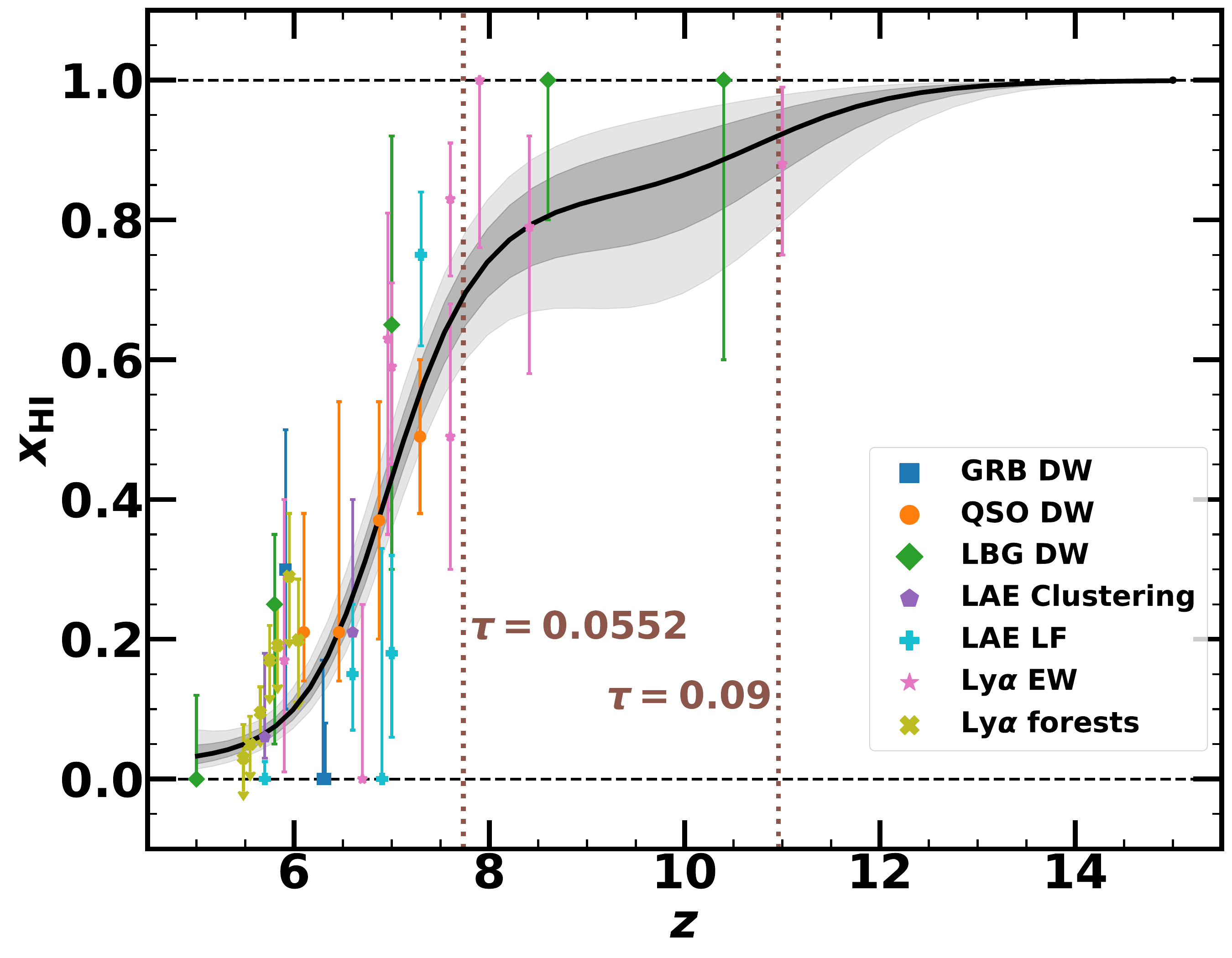}
    \caption{Cosmic reionization history reconstructed via Gaussian process regression. The black line represents the median reionization history, while the gray shaded regions show the 68\% (dark gray) and 95\% (light gray) posterior ranges. We also present $\xHI$ estimates based on various methods used in this work: Ly$\alpha$ damping-wing absorption of GRBs (blue squares; \citealt{totani06, totani14, fausey25}), QSOs (orange circles; \citealt{greig22, durovcikova24}), LBGs (green diamonds; \citealt{umeda25b}), LAE clustering (purple pentagons; \citealt{umeda25a}), Ly$\alpha$ luminosity function (cyan pluses; \citealt{wold22, umeda25a}), Ly$\alpha$ EW distribution (pink stars; \citealt{mason19a, jung20, whitler20, bolan22, kageura25}), and Ly$\alpha$/$\beta$ forests (olive crosses; \citealt{zhu22, davies25}). \revise{Error bars indicate 1$\sigma$ uncertainties. One-sided constraints are shown as error bars anchored at $\xHI=0$ or $\xHI=1$, except for the Ly$\alpha$-forest constraints, which are shown as the quoted upper limits.} The brown dotted lines indicate the redshift corresponding to instantaneous reionization with $\tau=0.09$ and $\tau=0.0552$, for $\Omega_bh^2=0.02237$, $\Omega_ch^2=0.1200$, and $h=0.6736$. We note that the instantaneous reionization redshift $z=7.73$ for $\tau=0.0552$ differs from the midpoint of reionization $z_\mathrm{mid}=7.19$, since the actual reionization history is not instantaneous.}
    \label{fig:gp_crh}
\end{figure}

To derive the CMB optical depth $\tau$, the reionization history $\xHI(z)$ as a function of redshift is required.
To reconstruct the reionization history, we adopt Gaussian process regression.
Gaussian process regression is a non-parametric method for reconstructing a function from a finite set of data points without assuming any specific physical model, and it can be regarded as a generalization of binning \citep{lodha25}.
Using constraints on the neutral fraction $\xHI$ at various redshifts $\vb{z}=(z_1,z_2,...,z_n)$ (Section \ref{sec:crh}), $\xHI$ values at new redshift points $\vb{z}^\star=(z_1^\star,z_2^\star,...,z_{n'}^\star)$ can be obtained as a posterior distribution in a Bayesian framework.
Unlike approaches that construct a reionization model by solving an ionization differential equation from high to low redshift (or vice versa), our method infers the values of $\xHI$ at all redshift points simultaneously.
The Gaussian process prior further correlates nearby redshift points, favoring similar $\xHI$ values at similar redshifts and thereby yielding a smooth, continuous reionization history consistent with the data.
We calculate the posterior distribution $p(\xHI(\vb{z}^\star)|\{\xHI\})$ using Gaussian process regression, where $\{\xHI\}$ denotes the full set of constraints on $\xHI(\vb{z})$.
We use the Markov Chain Monte Carlo (MCMC) method with the \texttt{emcee} package \citep{foremanmackey13} to perform the regression, with 62 chains and $10^5$ steps.
See Appendix \ref{sec:gp} for details of the regression methodology.
For Gaussian process reconstruction of the reionization history, see also \revise{\citet{krishak21} and \citet{elbers25}}.
The reconstructed reionization history is shown in Figure \ref{fig:gp_crh}.
In the reconstructed history, reionization primarily occurs at $z\sim6-8$, which is consistent with previous astrophysical results \citep{naidu20, matthee22, kageura25, shimizu25}.
In comparison, under an instantaneous reionization scenario, an optical depth of $\tau\sim0.09$ implies that reionization occurs at $z\sim11$.
However, our results indicate that the neutral hydrogen fraction remains as high as $\xHI(z)\sim0.9$ at $z\sim11${\revise{, suggesting that such a high optical depth is unlikely within our Gaussian process reconstruction.}}
\revise{\citet{cain25} also estimated the duration of reionization implied by $\tau\sim0.09$ using Ly$\alpha$ forest constraints, and discussed a tension with the SPT limit from the patchy kinematic Sunyaev–Zel'dovich (pkSZ) effect.}

\section{Constraints on the base-$\Lambda$CDM model}\label{sec:lcdm}
\subsection{Likelihood and Methodology}\label{sec:method}
We first constrain cosmological parameters within $\Lambda$CDM cosmology.
In our analysis, we simultaneously fit five cosmological parameters and the reionization history using the CMB likelihoods and the Gaussian process regression results.
For the cosmological parameters, we include the baryon density $\omega_b=\Omega_bh^2$, cold dark matter density $\omega_c=\Omega_ch^2$, Hubble constant $H_0$, primordial curvature power spectrum amplitude $\ln(10^{10}A_s)$, and spectral index $n_s$ to fit the CMB power spectra.
The sum of neutrino masses is fixed at $\Sigma m_\nu=0.06$ eV.
For the reionization history, we use $\xHI$ values at ten redshift points, $\xHI(z_1), ..., \xHI(z_{10})$ as fitting parameters.
We set these redshift points at equal intervals in $\log(1+z)$-space, spanning $z_1=5$ to $z_{10}=15$.
We confirm that a 10-point sampling is sufficient for the integration in Equation (\ref{equ:tau}) to converge and that $\tau$ is calculated with an accuracy of $10^{-4}$, which is significantly smaller than the statistical error.
The posterior distribution of the cosmological parameters $\vb*{\theta}$ and reionization history $\xHI(\vb{z}^\star)$, given the CMB data $d_\mathrm{CMB}$ and reionization constraints $\{\xHI\}$, is expressed as
\revise{
\begin{align}\label{equ:post}
    \log{p}&(\vb*{\theta},\xHI(\vb{z}^\star)|d_\mathrm{CMB},\{\xHI\})\notag\\
    &=\log{p(\xHI(\vb{z}^\star)|\{\xHI\})}+\log \mathcal{L}_\mathrm{CMB}(\vb*{\theta},\xHI(\vb{z}^\star)|d_\mathrm{CMB})\notag\\
    &\quad+\log{p(\vb*{\theta})}+const.
\end{align}
}
\par
The first term represents the posterior distribution from the Gaussian process.
\revise{The second term $\mathcal{L}_\mathrm{CMB}(\vb*{\theta},\xHI(\vb{z}^\star)|d_\mathrm{CMB})$ denotes the CMB likelihood, constructed from the
\texttt{CamSpec}, \texttt{Commander}, and CMB-lensing likelihoods listed in Table \ref{tab:cmb}.}
\revise{The third term is the prior on the cosmological parameters.}
\revise{In Eq. (\ref{equ:post}), $\vb*{\theta}$ denotes the set of cosmological parameters $(\omega_b=\Omega_bh^2,\omega_c=\Omega_ch^2,H_0,\ln(10^{10}A_s),n_s)$. 
The quantity $\xHI(\vb{z}^\star)$ represents the neutral hydrogen fraction evaluated at the redshift
points $\vb{z}^\star=(z^\star_1,\ldots,z^\star_{10})$, while $\{x_{\rm HI}\}$ denotes the set of observational constraints on the
neutral fraction described in Section \ref{sec:crh}.}
In this framework, we directly fit the reionization history and do not treat $\tau$ as an explicit fitting parameter.
Instead, $\tau$ is derived from $\omega_b$, $\omega_c$, $H_0$, and the reionization history using the following formula:
\begin{align}\label{equ:tau}
    \tau=&\frac{3c\omega_bX_p\sigma_T\times100~\mathrm{km/s}}{8\pi Gm_p}\notag\\
    &\times\int\dd{z}\frac{(1-\xHI(z))(1+z)^2}{\sqrt{(\omega_b+\omega_c)((1+z)^3-1)+h^2}}\qty(1+\frac{\eta Y_p}{4X_p}),
\end{align}
where $c$ is the speed of light, $X_p$ is the hydrogen fraction, $Y_p$ is the helium fraction, $\sigma_T$ is the Thomson-scattering cross section, $G$ is the gravitational constant, and $m_p$ is the proton mass.
The parameter $\eta$ accounts for additional electrons supplied by helium reionization.
Following \citet{planck18_6}, we assume that the first helium reionization occurs contemporaneously with hydrogen reionization, and the second helium reionization takes place at $z=3.5$ (i.e., $\eta=1$ at $z>3.5$ and $\eta=2$ at $z\leq3.5$).
\begin{deluxetable}{c|ccc}
    \tablecolumns{4}
    \tablewidth{\linewidth}
    \tabletypesize{\scriptsize}
    \tablecaption{Parameter means and $68\%$ credible intervals for the base-$\Lambda$CDM model from CMB and reionization history.
    \label{tab:lcdm}}
    \tablehead{
    Parameter & CMB (w/o lowE) & CMB (w/ lowE) & CMB (w/o lowE)+$\xHI(z)$}
    \startdata
    $\tau$ & $0.072\pm 0.016$ & $0.0539\pm 0.0073$ & $0.0552^{+0.0019}_{-0.0026}$\\
    $\ln(10^{10}A_s)$ & $3.074\pm 0.028$ & $3.043\pm 0.013$ & $3.0451^{+0.0055}_{-0.0062}$\\
    $n_s$ & $0.9666\pm 0.0047$ & $0.9634\pm 0.0040$ & $0.9638\pm 0.0038$\\
    $\Omega_bh^2$ & $0.02226\pm 0.00015$ & $0.02218\pm 0.00013$ & $0.02218\pm 0.00013$\\
    $\Omega_ch^2$ & $0.1190\pm 0.0013$ & $0.1200\pm 0.0011$ & $0.11993\pm 0.00099$\\
    $H_0~\mathrm{[km~s^{-1}~Mpc^{-1}]}$ & $67.61\pm 0.60$ & $67.14\pm 0.47$ & $67.17\pm 0.44$\\
    $\Omega_m$ & $0.3104\pm 0.0082$ & $0.3169\pm 0.0065$ & $0.3165\pm 0.0061$\\
    $\sigma_8$ & $0.8205\pm 0.0092$ & $0.8106\pm 0.0051$ & $0.8115\pm 0.0036$\\
    $S_8$ & $0.834\pm 0.010$ & $0.833\pm 0.011$ & $0.833\pm 0.011$\\
    $H_0r_d~\mathrm{[km~s^{-1}]}$ & $9891\pm 81$ & $9973\pm 100$ & $9896\pm 76$
    \enddata
\end{deluxetable}
\revise{
The CMB likelihood term in Eq. (\ref{equ:post}), $\mathcal{L}_\mathrm{CMB}(\vb*{\theta},\xHI(\vb{z}^\star)|d_\mathrm{CMB})$, is evaluated using the CMB datasets summarized in Table \ref{tab:cmb}.
For each MCMC sample, we first derive $\tau$ from $\vb*{\theta}$ and $\xHI(\vb{z}^\star)$ using Eq. (\ref{equ:tau}).
We then compute the theoretical CMB angular power spectra with \texttt{CAMB} \citep{lewis00, howlett12} and evaluate the likelihoods with \texttt{cobaya} \citep{torrado19, torrado21}.
The prior term in Eq. (\ref{equ:post}) is taken to be flat over the parameter ranges adopted by \citet{planck13_16} and \citet{planck18_6}.
Although our primary results are based on CMB (w/o lowE) and reionization history $\xHI(z)$, we also derive results using CMB (w/o lowE) alone for comparison.}
The posterior distributions are sampled using the MCMC method with \texttt{emcee} with $10^4$ steps and $2\times(\mathrm{number~of~parameters})$ walkers.

\subsection{Cosmological Parameters}\label{sec:cosmological_parameters}
\begin{figure*}
    \centering
    \includegraphics[width=\linewidth]{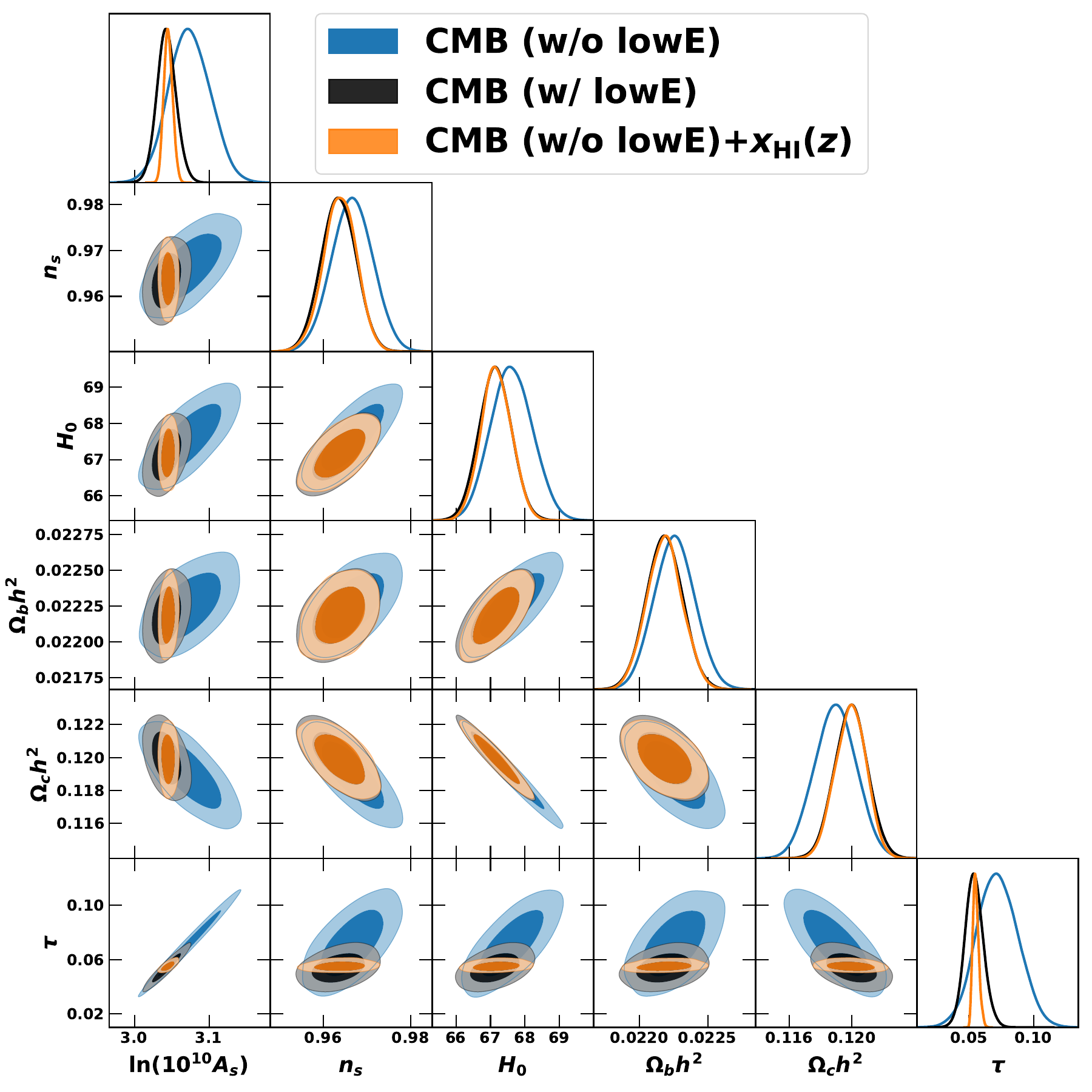}
    \caption{Constraints on cosmological parameters and $\tau$ for the $\Lambda$CDM model. The orange contours represent constraints derived from CMB Planck TTTEEE and Planck+ACT lensing and cosmic reionization history, excluding low-$l$ EE. The blue and black contours indicate constraints obtained from CMB (w/o lowE) and CMB (w/ lowE), respectively. The dark and light contours show the 68\% and 95\% regions, respectively.}
    \label{fig:contour}
\end{figure*}
\begin{figure}
    \centering
    \includegraphics[width=\linewidth]{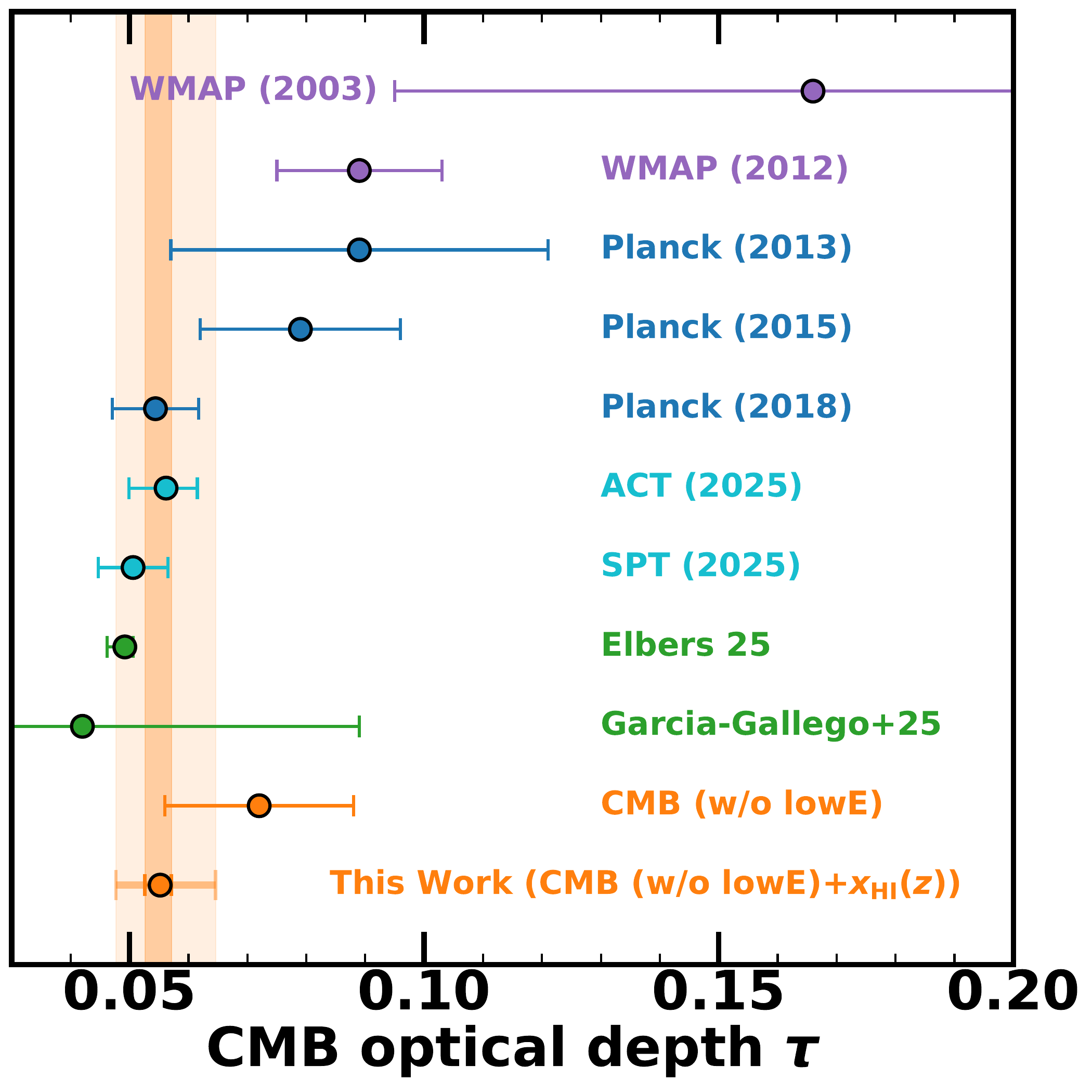}
    \caption{Comparison of our $\tau$ values with those from previous studies. The orange circles show our $\tau$ values from CMB (w/o lowE) and CMB (w/o lowE)+$\xHI(z)$. \revise{The dark and light orange shaded regions indicate the statistical uncertainty and the uncertainty including the possible absorption-model systematic, respectively.} The purple and blue points indicate $\tau$ values from WMAP \citep{wmap03, wmap12} and Planck \citep{planck13_16, planck15_13, planck18_6}, respectively. The cyan circles show results from ACT \citep{act25} and SPT \citep{spt25} combined with Planck data. The green circles represent $\tau$ constraints from \citet{elbers25} based on reionization history, BAO, and BBN, and from \citet{garciagallego25} based on the Ly$\alpha$ forest.}
    \label{fig:tau_ref}
\end{figure}
In Figure \ref{fig:contour} and Table \ref{tab:lcdm} we show the posterior distributions of the cosmological parameters and $\tau$.
The MCMC samples are analyzed and plotted using \texttt{GetDist} \citep{lewis25}.
For comparison, we also show the results from CMB (w/ lowE) based on the publicly available MCMC chains provided by the DESI Collaboration\footnote{\url{https://data.desi.lbl.gov/public/papers/y3/bao-cosmo-params/README.html}}.
By combining reionization history data with CMB (w/o lowE), we obtain a stringent constraint on the optical depth,
\begin{align}
    \tau=0.0552^{+0.0019}_{-0.0026}\quad(\mathrm{CMB~(w/o~lowE)}+\xHI(z)).
\end{align}
This result is consistent with the constraint from CMB large-scale E-mode polarization,
\begin{align}
    \tau=0.0539\pm0.0072\quad(\mathrm{CMB~(w/~lowE)}).
\end{align}
Without reionization history, we obtain
\begin{align}
    \tau=0.072\pm 0.016\quad(\mathrm{CMB~(w/o~lowE)}).
\end{align}
Incorporating $\xHI(z)$ data improves the estimate of $\tau$ by a factor of $7$.\par
\revise{We compare our constraint on $\tau$ with literature values in Figure \ref{fig:tau_ref}.
Our result is broadly consistent with recent astrophysical constraints on $\tau$ that do not rely on large-scale CMB E-mode polarization. 
\citet{elbers25} obtained $\tau=0.0492^{+0.0014}_{-0.0030}$ by combining reionization history with BAO and BBN.
If only the statistical uncertainty of our fiducial analysis is considered, our central value is slightly higher than that of \citet{elbers25}.
However, once the possible absorption-model systematic uncertainty is included, our constraint becomes $\tau=0.0552^{+0.0019}_{-0.0026}(\mathrm{stat.})^{+0.0075}_{-0.0049}(\mathrm{sys.})$, making the two results consistent within the quoted uncertainties (Section \ref{sec:sys}).
Our result is also consistent with the Ly$\alpha$-forest-based constraint of \citet{garciagallego25}, although their allowed range is substantially broader.
This larger uncertainty is expected because their method infers the reionization history indirectly through the thermal memory of the IGM from the Ly$\alpha$ forest alone, and is therefore sensitive to degeneracies among the reionization timing, duration, and thermal history.
In contrast, our analysis reconstructs $\xHI(z)$ over $z\simeq5$--14 using multiple probes of the neutral fraction and combines this reionization history with small-scale CMB and CMB-lensing data, which solves the degeneracy between $\tau$ and other cosmological parameters.}\par
Because $\tau$ is degenerate with other cosmological parameters, constraints on other cosmological parameters also improve when $\xHI(z)$ data are included.
In particular, the constraint on the primordial power spectrum amplitude $A_s$ significantly improves when $\xHI(z)$ is included, since the amplitude of the small-scale CMB power spectrum scales with $A_se^{-2\tau}$.
We obtain
\begin{align}
    &\ln(10^{10}A_s)=3.0451^{+0.0055}_{-0.0062}\notag\\
    &\qquad(\mathrm{CMB~(w/o~lowE)}+\xHI(z)),
\end{align}
which is consistent with the lowE result, $\ln(10^{10}A_s)=3.043\pm0.013$ (CMB (w/ lowE)).
By solving the degeneracy between $A_s$ and $\tau$, our constraint on $A_s$ is improved by a factor of $5$ relative to that from CMB (w/o lowE) alone, $\ln(10^{10}A_s)=3.074\pm 0.028$.\par
Because increases in $\tau$ and $n_s$ both suppress power at high-$l$, these two parameters are also correlated \citep{mcdonough25}.
We obtain constraints on $n_s$ as
\begin{align}
    n_s=0.9638\pm 0.0038\quad(\mathrm{CMB~(w/o~lowE)}+\xHI(z)),
\end{align}
which is consistent with the lowE result, $n_s=0.9634\pm0.0040$ (CMB (w/ lowE)).
The value of $n_s$ from CMB (w/o lowE)+$\xHI(z)$ is in agreement with the Higgs, Starobinsky, and exponential $\alpha$-attractor inflation models \citep{kallosh25}.

\subsection{Comparison with BAO Measurements}\label{sec:results_bao}
\begin{figure*}
    \centering
    \includegraphics[width=\linewidth]{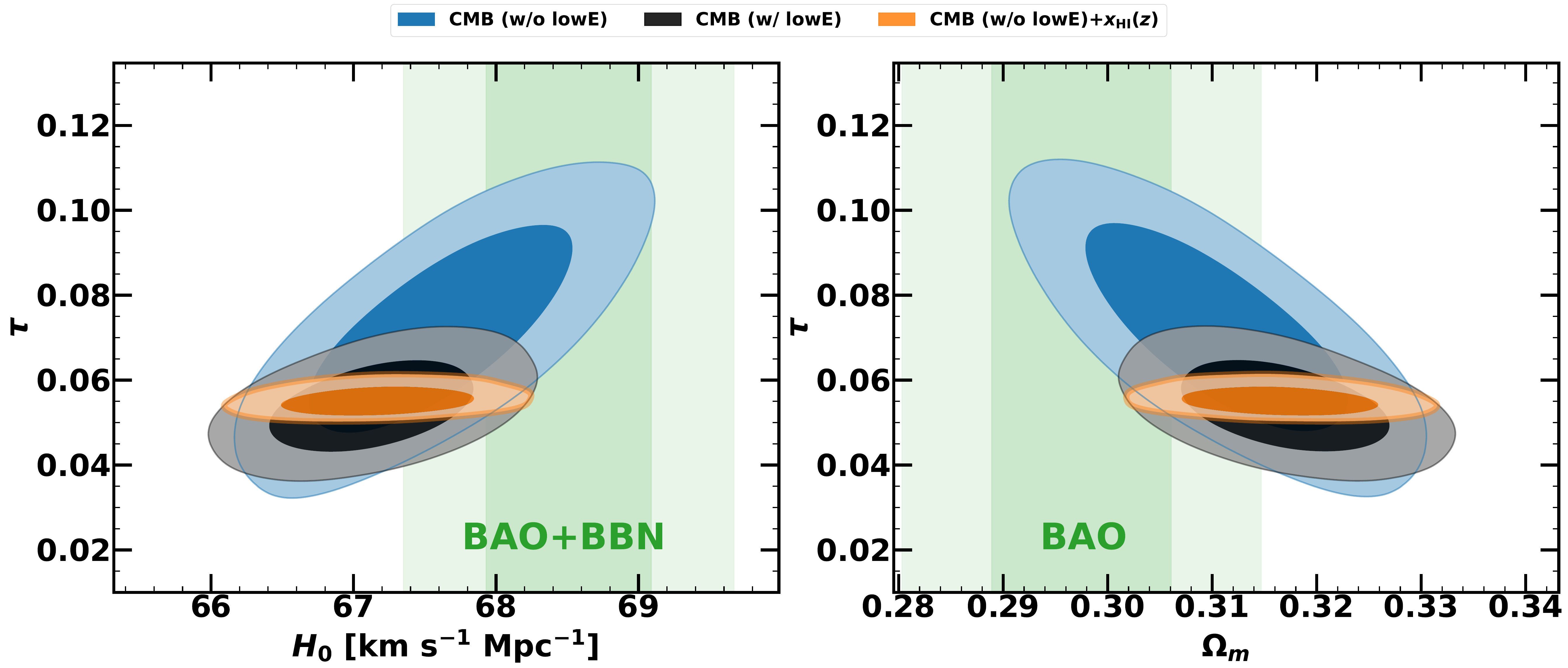}
    \caption{Constraints on $H_0$ and $\Omega_m$ for the $\Lambda$CDM model. The blue, black, and orange contours show results based on CMB (w/o lowE), CMB (w/ lowE), and CMB (w/o lowE)+$\xHI(z)$, respectively. The green region for $\Omega_m$ indicates constraints from DESI DR2 BAO measurements, while the green region for $H_0$ shows results from DESI DR2 BAO data with a BBN prior on $\omega_b$. Contours show the 68\% and 95\% regions.}
    \label{fig:H0_omegam_tau}
\end{figure*}
\begin{figure}
    \centering
    \includegraphics[width=\linewidth]{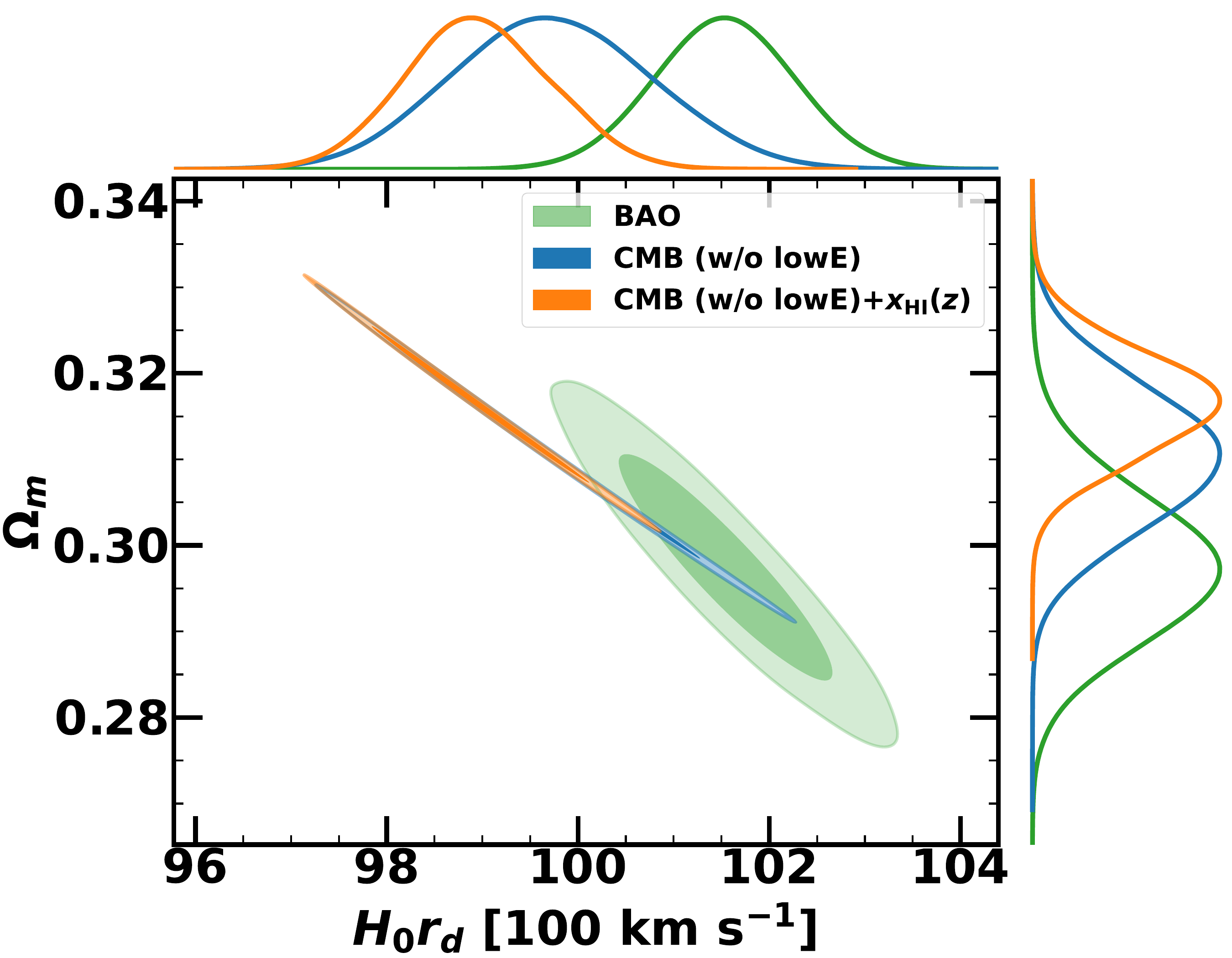}
    \caption{Constraints on $H_0r_d$ and $\Omega_m$ for the $\Lambda$CDM model from DESI DR2 BAO (green), CMB (w/o lowE) (blue), and CMB (w/o lowE)+$\xHI(z)$ (orange). The result including reionization history in this work is consistent with the CMB (w/ lowE) constraint and is in $2.4\sigma$ tension with the DESI DR2 BAO measurements. Contours show the 68\% and 95\% regions.}
    \label{fig:H0rd_omegam}
\end{figure}
As shown in Figure \ref{fig:contour}, $H_0$, $\Omega_ch^2$, and $\tau$ are correlated with each other.
Consequently, the precise constraint on $\tau$ leads to improved constraints on the Hubble constant $H_0$, matter density $\Omega_m$, and sound horizon scale $r_d$ as follows:
\begin{align}
    &H_0=67.17\pm 0.44~\mathrm{[km~s^{-1}~Mpc^{-1}]}\notag\\
    &\qquad(\mathrm{CMB~(w/o~lowE)}+\xHI(z))
\end{align}
\begin{align}
    \Omega_m=0.3165\pm 0.0061\quad(\mathrm{CMB~(w/o~lowE)}+\xHI(z))
\end{align}
\begin{align}
    &H_0r_d=9896\pm76~\mathrm{[km~s^{-1}]}\notag\\
    &\qquad(\mathrm{CMB~(w/o~lowE)}+\xHI(z)).
\end{align}
We show these constraints in Figure \ref{fig:H0_omegam_tau}.
We compare our results with those from BAO measurements.
DESI DR2 finds $\Omega_m=0.2975\pm0.0086$, and $H_0=68.51\pm0.58~\mathrm{[km~s^{-1}~Mpc^{-1}]}$ with a BBN prior on $\omega_b$ \citep{desidr2_2}.
These $\Omega_m$ and $H_0$ values are fully consistent with those obtained from CMB (w/o lowE) alone.
In contrast, the results from CMB (w/o lowE)+$\xHI(z)$ exhibit mild discrepancies with the BAO constraints, with a $1.8\sigma$ difference for both $\Omega_m$ and $H_0$.
\revise{For completeness, we also show the same comparison in the $\Omega_\Lambda$--$\tau$ plane in Appendix~\ref{sec:omegal}.}
To further examine whether the results based on CMB (w/o lowE)+$\xHI(z)$ are in tension with DESI DR2 BAO measurements, we evaluate the discrepancy in the $H_0r_d-\Omega_m$ plane (Figure \ref{fig:H0rd_omegam}).
We use the MCMC chains of the posterior distributions and covariance matrices of cosmological parameters that are publicly provided by the DESI Collaboration\footnote{\url{https://data.desi.lbl.gov/public/papers/y3/bao-cosmo-params/README.html}}.
Following \citet{desidr2_2}, we compute the relative $\chi^2$ between CMB (w/o lowE)+$\xHI(z)$ and BAO as
\begin{align}
    \chi^2=(\vb{p}_C-\vb{p}_B)^T(\mathrm{Cov}_C+\mathrm{Cov}_B)^{-1}(\vb{p}_C-\vb{p}_B),
\end{align}
where $\vb{p}_C$ ($\vb{p}_B$) and $\mathrm{Cov}_C$ ($\mathrm{Cov}_B$) represent the $(H_0r_d, \Omega_m)$ values and covariance matrices from CMB (w/o lowE)+$\xHI(z)$ (BAO), respectively.
We convert this $\chi^2$ value to a probability-to-exceed (PTE) value and find a $2.4\sigma$ tension between CMB (w/o lowE)+$\xHI(z)$ and BAO measurements.
This result is consistent with the $2.3\sigma$ discrepancy between BAO and CMB including large-scale E-mode polarization reported by the DESI Collaboration \citep{desidr2_2}.
Although the BAO measurements are consistent with CMB excluding large-scale polarization measurements that may be affected by instrumental and foreground systematics, the inclusion of the reionization history helps to solve the degeneracies between $\tau$ and $\Omega_m$, as well as between $\tau$ and $H_0r_d$.
This confirms the CMB-BAO tension without relying on large-scale E-mode polarization measurements.
\revise{For comparison, we also constrain cosmological parameters using both large-scale E-mode polarization and the reionization history, and find that the results are nearly identical to those obtained without lowE.
For details, see Appendix \ref{sec:lowE_xHI}.}
Since the impact of the $\Omega_m$-$\tau$ degeneracy is largely removed, future BAO measurements will be required to investigate this tension with higher statistical precision.\par
\revise{
We also examine how the possible systematic uncertainty in $\tau$ discussed in Section \ref{sec:sys} propagates into the parameters shown in this section.
When the absorption-model systematic in $\xHI(z)$ is included, we obtain
\begin{align}
\Omega_m &= 0.3165 \pm 0.0061\,({\rm stat.})^{+0.0009}_{-0.0007}\,({\rm sys.}), \\
H_0 &= 67.17 \pm 0.44\,({\rm stat.})^{+0.04}_{-0.07}\,({\rm sys.})
\,{\rm km\,s^{-1}\,Mpc^{-1}} .
\end{align}
These systematic shifts are much smaller than the statistical uncertainties.
This is because the inclusion of the reionization-history information largely resolves the $\Omega_m$--$\tau$ and $H_0$--$\tau$ degeneracies, so that the marginal uncertainties in $\Omega_m$ and $H_0$ are no longer dominated by their covariance with $\tau$.
Therefore, even under this conservative treatment of the $\tau$ systematic, our comparison with BAO measurements and the inferred CMB--BAO tension are essentially unchanged.
}

\subsection{The Fluctuation Amplitude}
\begin{figure*}
    \centering
    \includegraphics[width=\linewidth]{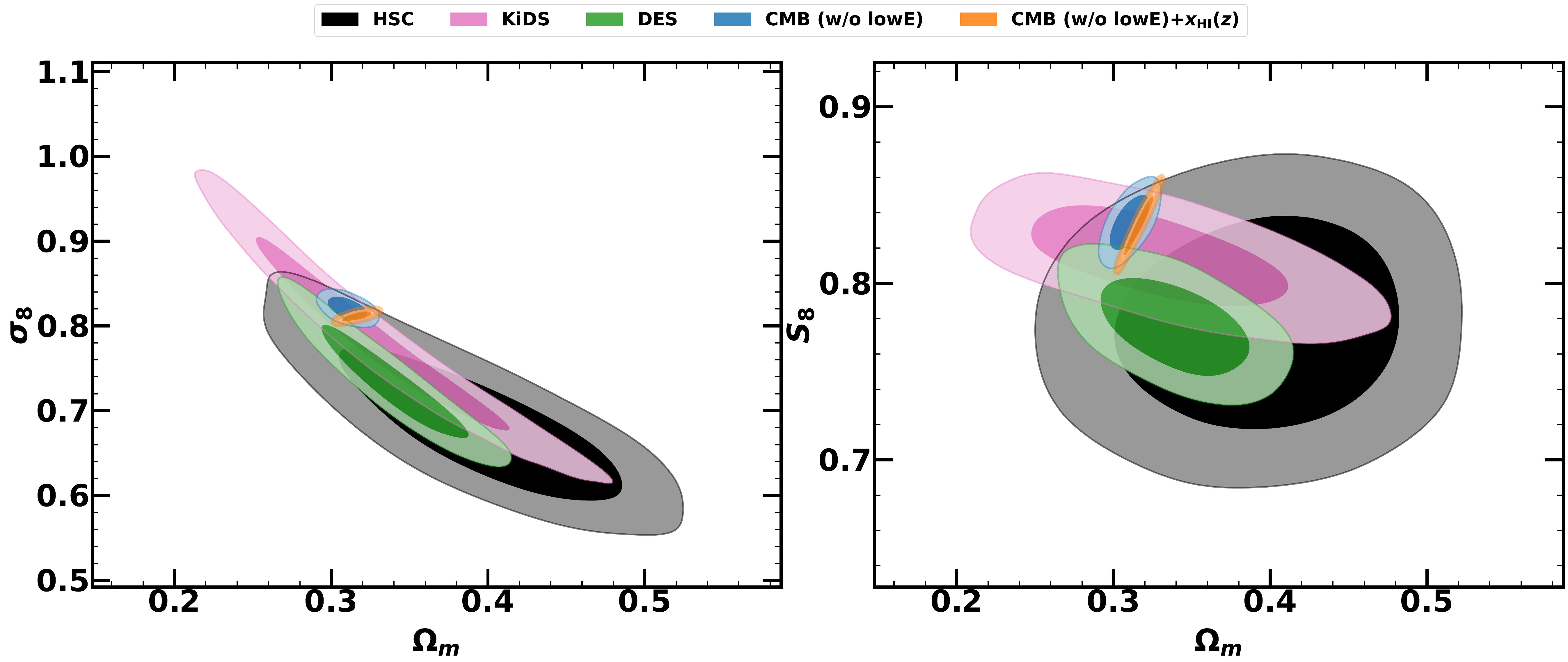}
    \caption{Constraints on $\sigma_8$ and $S_8$ for the $\Lambda$CDM model from HSC-Y3 (black), KiDS-Legacy (pink), DES Y3 (green), CMB (w/o lowE) (blue), and CMB (w/o lowE)+$\xHI(z)$ (orange). The $S_8$ value obtained from CMB (w/o lowE)+$\xHI(z)$ is $2.7\sigma$ higher than the DES Y3 constraint, while remaining consistent with the HSC-Y3, KiDS-Legacy, and CMB (w/ lowE) results. Contours show the 68\% and 95\% regions.}
    \label{fig:sigma8}
\end{figure*}
As discussed in Section \ref{sec:cosmological_parameters}, the CMB optical depth $\tau$ and primordial power spectrum amplitude $A_s$ are strongly correlated.
A precise determination of $\tau$ using reionization history thus leads to a simultaneous improvement in the constraint on $A_s$.
Under the $\Lambda$CDM model, $A_s$ can be converted into the fluctuation amplitude at $z=0$, $\sigma_8$.
We compute $\sigma_8$ using the \texttt{CAMB} code and obtain
\begin{align}
    \sigma_8=0.8115\pm 0.0036\quad(\mathrm{CMB~(w/o~lowE)}+\xHI(z)).
\end{align}
Our result is consistent with the CMB constraint including large-scale E-mode polarization, $\sigma_8=0.8106\pm0.0051$ (CMB (w/ lowE)).
Our constraint is $\sim3$ times tighter than that obtained using CMB (w/o lowE) alone, $\sigma_8=0.8205\pm0.0092$ (CMB (w/o lowE)).\par
Galaxy clustering measurements provide an independent constraint on the fluctuation amplitude.
We compare our results with data from the Dark Energy Survey (DES) Y3.
For this comparison, we use the cosmology chains for the fiducial $\Lambda$CDM $3\times2$pt likelihood, which combines two-point correlation functions of cosmic shear, galaxy clustering, and the cross-correlation of source galaxy shear with lens galaxy positions, galaxy–galaxy lensing \citep{des05, flaugher15, des16, abbott22a, abbott22b, amon22, secco22}\footnote{\url{https://des.ncsa.illinois.edu/releases/y3a2/Y3key-products}}.\par
In Figure \ref{fig:sigma8}, we present our results together with the DES Y3 constraints.
We calculate $S_8=\sigma_8\sqrt{\Omega_m/0.3}$ and find
\begin{align}
    S_8=0.833\pm 0.011\quad(\mathrm{CMB~(w/o~lowE)}+\xHI(z)).
\end{align}
Our $S_8$ value is consistent with the lowE result of $S_8=0.833\pm0.011$ (CMB (w/ lowE)) and is larger than the DES Y3 measurement, $S_8=0.776\pm0.018$, by $2.7\sigma$.
In contrast, our result is consistent with the $3\times2$pt analysis of Year 3 Hyper Suprime-Cam (HSC-Y3) and the cosmic shear constraints from the Kilo-Degree Survey (KiDS), which report $S_8=0.775_{-0.038}^{+0.043}$ \citep{miyatake23, more23, sugiyama23}\footnote{Cosmological chains are publicly available at \url{https://hsc-release.mtk.nao.ac.jp/doc/index.php/s19a-shape-catalog-pdr3/}.} and $S_8=0.815_{-0.021}^{+0.016}$ \citep{wright24, wright25}\footnote{Cosmological chains are publicly available at \url{https://kids.strw.leidenuniv.nl/}.}, respectively.
Further investigations are needed to clarify the presence of the $S_8$ tension.

\subsection{Possible Systematics in the Determination of $\tau$}\label{sec:sys}
Our reconstruction of the reionization history relies on measurements of $\xHI$ inferred from Ly$\alpha$ damping-wing absorption in the spectra of galaxies and QSOs.
A key caveat is that such inferences are intrinsically model dependent.
The damping-wing strength is sensitive not only to the mean neutral fraction but also to the characteristic size distribution of ionized bubbles.
Therefore, assumptions about the IGM topology can bias the inferred $\xHI$.
\citet{kageura25} quantified this effect by estimating $\xHI$ using a standard model and an extreme scenario with very large ionized bubbles.
They found that the inferred $\xHI$ can shift by up to $\sim7\%$ between these models.
For individual data points, this uncertainty is smaller than the statistical errors.
However, a coherent bias of comparable magnitude affecting all measurements could propagate non-negligibly into the determination of $\tau$.\par
To account for this potential systematics, we propagate a $\sim7\%$ modeling uncertainty in the $\xHI$ values and obtain $\tau=0.0552_{-0.0049}^{+0.0075}$ (systematics).
Under this conservative treatment, the systematic uncertainty exceeds the statistical uncertainty while remaining comparable to the uncertainty in the large-scale CMB polarization constraints ($\tau=0.0539\pm0.0072$; CMB (w/ lowE)).
Therefore, our discussion of the CMB-BAO tension is not qualitatively affected by this possible modeling systematics.\par
We note that there are potential systematic biases that are not included in the discussion above.
For example, the intrinsic spectra prior to IGM absorption must be assumed to evaluate the Ly$\alpha$ damping-wing absorption.
In many studies, galaxy/QSO spectra at $z\lesssim5$ are used to model the intrinsic spectra at the Epoch of Reionization (EoR).
However, the intrinsic properties of galaxies and QSOs at the EoR may differ from those at lower redshifts.
For example, \citet{rinaldi23} found the typical H$\alpha$ EW of galaxies evolve as $\propto(1+z)^{2.1}$, potentially suggesting that Ly$\alpha$ and UV continuum emission also evolve with redshift.\par
The resolution of simulations can also affect the inferred value of $\xHI$.
To account for the inhomogeneous ionization state of the IGM during the EoR, many studies estimate $\xHI$ by comparing observed galaxy and QSO properties with IGM simulations.
However, capturing the inhomogeneity of reionization requires simulation volumes on Gpc scales, larger than typical ionized bubbles, while small-scale physics is also known to play a major role in regulating the progress of reionization and the absorption of background radiation.
For example, \citet{cain24} showed that calculations of Ly$\alpha$ forest power spectrum have not yet converged even in simulations with a high spatial resolution of $\sim2$ kpc per cell, and it remains unclear which observables are affected by simulation resolution and to what extent.\par
Quantitative assessments of these potential sources of systematic uncertainty therefore require further investigation in future work.

\section{Dynamical Dark Energy}\label{sec:dde}
\begin{deluxetable}{c|cccc}
    \tablecolumns{5}
    \tablewidth{\linewidth}
    \tabletypesize{\scriptsize}
    \tablecaption{Parameter means and $68\%$ credible intervals for the $w_0w_a$ model from CMB, BAO, and reionization history.
    \label{tab:dde}}
    \tablehead{
    Parameter & CMB (w/o lowE) & CMB (w/o lowE)+BAO & CMB (w/ lowE)+BAO & CMB (w/o lowE)+BAO+$\xHI(z)$}
    \startdata
    $w_0$ & $-1.12^{+0.43}_{-0.48}$ & $-0.51^{+0.21}_{-0.24}$ & $-0.42\pm 0.21$ & $-0.55^{+0.17}_{-0.26}$\\
    $w_a$ & $-0.5^{+1.4}_{-1.2}$ & $-1.44^{+0.73}_{-0.58}$ & $-1.75\pm 0.58$ & $-1.40^{+0.70}_{-0.47}$\\
    $\tau$ & $0.065\pm 0.018$ & $0.065^{+0.015}_{-0.017}$ & $0.0527\pm 0.0071$ & $0.0556^{+0.0018}_{-0.0027}$\\
    $\ln(10^{10}A_s)$ & $3.060\pm 0.033$ & $3.060^{+0.026}_{-0.029}$ & $3.038\pm 0.013$ & $3.0437\pm 0.0060$\\
    $n_s$ & $0.9658\pm 0.0047$ & $0.9662^{+0.0037}_{-0.0042}$ & $0.9644\pm 0.0037$ & $0.9649\pm 0.0034$\\
    $\Omega_bh^2$ & $0.02226\pm 0.00015$ & $0.02226\pm 0.00014$ & $0.02221\pm 0.00012$ & $0.02222\pm 0.00013$\\
    $\Omega_ch^2$ & $0.1191\pm 0.0014$ & $0.1190\pm 0.0011$ & $0.11960\pm 0.00083$ & $0.11940\pm 0.00077$\\
    $H_0~\mathrm{[km~s^{-1}~Mpc^{-1}]}$ & $75.2^{+7.2}_{-13}$ & $64.3\pm 1.9$ & $63.6^{+1.7}_{-2.1}$ & $64.8^{+2.4}_{-1.6}$\\
    $\Omega_m$ & $0.264^{+0.065}_{-0.078}$ & $0.344^{+0.020}_{-0.023}$ & $0.353\pm 0.021$ & $0.339^{+0.016}_{-0.027}$\\
    $\sigma_8$ & $0.881^{+0.063}_{-0.10}$ & $0.791\pm 0.018$ & $0.781^{+0.015}_{-0.018}$ & $0.792^{+0.020}_{-0.014}$\\
    $S_8$ & $0.810^{+0.045}_{-0.029}$ & $0.846\pm 0.013$ & $0.846^{+0.013}_{-0.012}$ & $0.841\pm 0.012$\\
    $H_0r_d~\mathrm{[km~s^{-1}]}$ & $11090^{+1000}_{-1900}$ & $9484\pm 290$ & $9378^{+250}_{-310}$ & $9558^{+360}_{-240}$
    \enddata
\end{deluxetable}
In Section \ref{sec:results_bao}, we show that our CMB (w/o lowE)+$\xHI(z)$ results for the ($H_0r_d$-$\Omega_m$) parameters are in $2.4\sigma$ tension with DESI DR2 BAO.
This tension is consistent with the $2.3\sigma$ CMB-BAO tension including E-mode polarization \citep{desidr2_2} and may suggest cosmology beyond $\Lambda$CDM, although further investigations are needed to confirm the tension.
One possible scenario to resolve this tension is the $w_0w_a$ (dynamical dark energy) model \citep{desidr2_2}.
In this section, we constrain cosmological parameters for the dynamical dark energy model with the equation-of-state parameter,
\begin{align}
    w=w_0+(1-a)w_a,
\end{align}
where $a$ is the scale factor, and $w_0$ and $w_a$ are constant parameters.
In $\Lambda$CDM cosmology, $w_0=-1$ and $w_a=0$.
Here, we vary these two parameters and adopt flat priors for $w_0$ and $w_a$ with ranges of $-3<w_0<1$ and $-3<w_a<2$, following \citet{desidr2_2}.
We use three datasets to constrain the dynamical dark energy model: (i) CMB (w/o lowE), (ii) CMB (w/o lowE) and DESI DR2 BAO, and (iii) CMB (w/o lowE), DESI DR2 BAO, and reionization history $\xHI(z)$.
We follow the method in Section \ref{sec:method} and obtain constraints on cosmological parameters (Table \ref{tab:dde}).
For comparison, we also show results from CMB (w/ lowE)+BAO based on the publicly available MCMC chains provided by the DESI Collaboration.\par
\begin{figure}
    \centering
    \includegraphics[width=\linewidth]{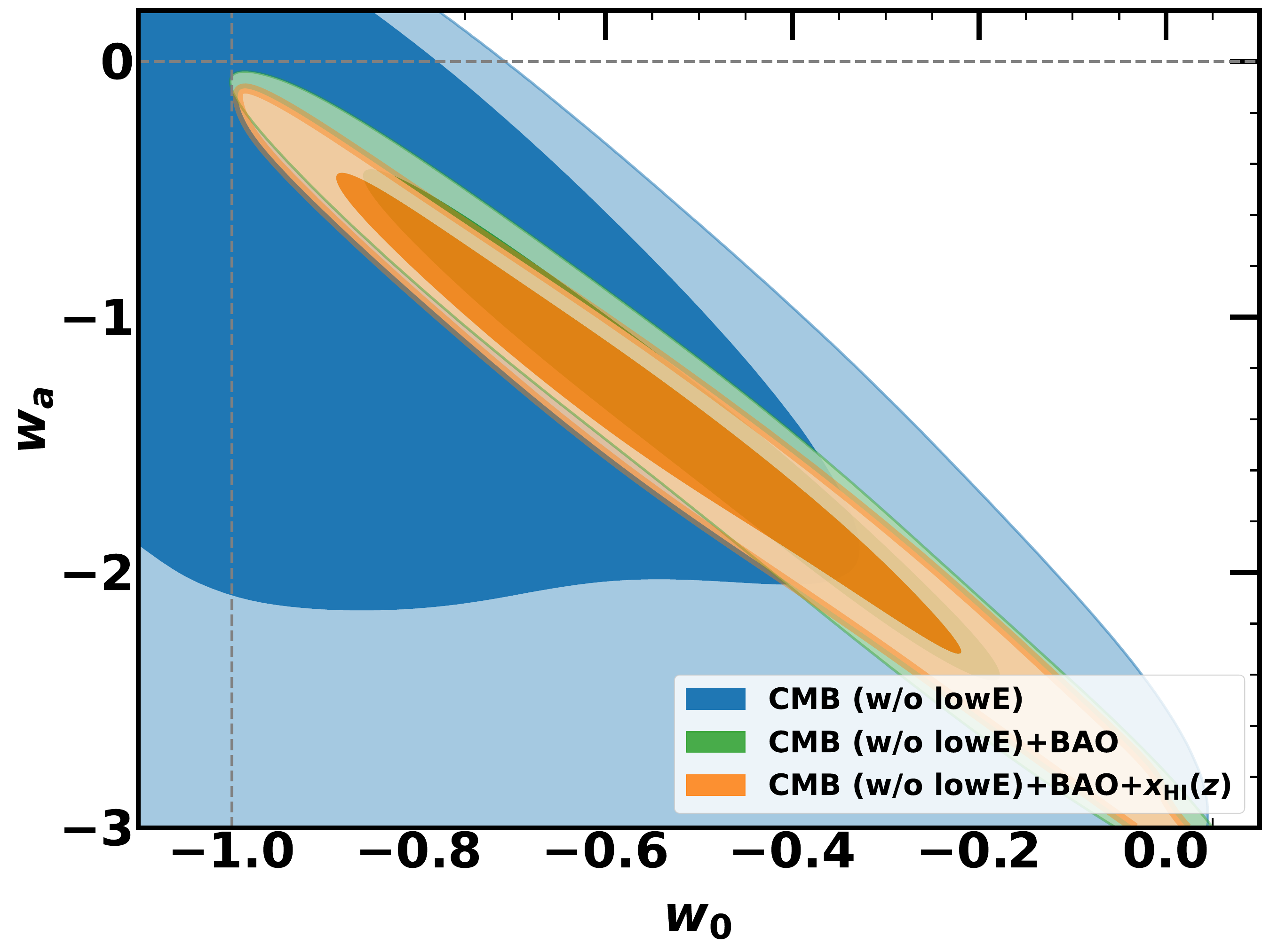}
    \caption{Marginalized posterior distributions of the ($w_0$-$w_a$) parameters from CMB (w/o lowE) (blue), +BAO (green), and +$\xHI(z)$ (orange). The dashed lines indicate $w_0=-1$ and $w_a=0$. The $\Lambda$CDM model is disfavored at $2.1\sigma$ significance by CMB (w/o lowE)+BAO+$\xHI(z)$, consistent with the CMB (w/ lowE)+BAO result reported by \citet{desidr2_2}. Contours show the $68\%$ and $95\%$ regions.}
    \label{fig:w0wa}
\end{figure}

The derived equation-of-state parameters are
\begin{align}
    &w_0=-0.55^{+0.17}_{-0.26}\quad w_a=-1.40^{+0.70}_{-0.47}\notag\\
    &\qquad(\mathrm{CMB~(w/o~lowE)+BAO}+\xHI(z)).
\end{align}
In Figure \ref{fig:w0wa}, we show the marginalized posterior distributions of the ($w_0$-$w_a$) parameters.
From CMB~(w/o~lowE) measurements, the ($w_0$-$w_a$) values are consistent with the $\Lambda$CDM model.
However, when BAO and reionization history $\xHI(z)$ are included, the posterior distribution deviates from the $\Lambda$CDM values of $(w_0,w_a)=(-1,0)$.
Following \citet{desidr2_2}, we compute the $\chi^2$ value between the best-fit $(w_0,w_a)$ and $(-1,0)$.
With this procedure, the $\Lambda$CDM model is disfavored at $2.1\sigma$ significance by the CMB (w/o lowE)+BAO+$\xHI(z)$ data.
This result is consistent with the lowE result, $w_0=-0.42\pm0.21$ and $w_a=-1.75\pm0.58$ (CMB (w/ lowE)+BAO), and aligns with previous studies \citep{desidr2_2, elbers25}.
Although this result requires further validation, if confirmed, it may indicate that dark energy is not a cosmological constant but instead reflects physics beyond a simple cosmological constant, such as an axion-like field \citep{tada24}.

\section{Sum of Neutrino Masses}\label{sec:mnu}
\begin{deluxetable}{c|cccc}
    \tablecolumns{5}
    \tablewidth{\linewidth}
    \tabletypesize{\scriptsize}
    \tablecaption{Parameter means and $68\%$ credible intervals for the base-$\Lambda$CDM model+$\Sigma m_\nu$ from CMB, BAO, and reionization history. We quote $95\%$ upper limits for $\Sigma m_\nu$.
    \label{tab:mnu}}
    \tablehead{
    Parameter & CMB (w/o lowE) & CMB (w/o lowE)+BAO & CMB (w/ lowE)+BAO & CMB (w/o lowE)+BAO+$\xHI(z)$}
    \startdata
    $\Sigma m_\nu~\mathrm{[eV]}$ & $< 0.445$ & $< 0.121$ & $< 0.0649$ & $< 0.0550$\\
    $\tau$ & $0.084^{+0.018}_{-0.022}$ & $0.083\pm 0.015$ & $0.0582^{+0.0065}_{-0.0075}$ & $0.0572^{+0.0021}_{-0.0042}$\\
    $\ln(10^{10}A_s)$ & $3.098^{+0.034}_{-0.041}$ & $3.092\pm 0.026$ & $3.049^{+0.012}_{-0.014}$ & $3.0465^{+0.0060}_{-0.0084}$\\
    $n_s$ & $0.9666\pm 0.0046$ & $0.9705^{+0.0033}_{-0.0039}$ & $0.9683\pm 0.0034$ & $0.9680\pm 0.0036$\\
    $\Omega_bh^2$ & $0.02225\pm 0.00015$ & $0.02237\pm 0.00013$ & $0.02232\pm 0.00012$ & $0.02233\pm 0.00012$\\
    $\Omega_ch^2$ & $0.1190\pm 0.0013$ & $0.11744^{+0.00085}_{-0.00071}$ & $0.11804\pm 0.00064$ & $0.11807\pm 0.00063$\\
    $H_0~\mathrm{[km~s^{-1}~Mpc^{-1}]}$ & $66.5^{+1.7}_{-1.0}$ & $68.49\pm 0.32$ & $68.37\pm 0.29$ & $68.40\pm 0.29$\\
    $\Omega_m$ & $0.325^{+0.012}_{-0.023}$ & $0.2991\pm 0.0040$ & $0.3009\pm 0.0037$ & $0.3006^{+0.0034}_{-0.0037}$\\
    $\sigma_8$ & $0.803^{+0.027}_{-0.015}$ & $0.8284\pm 0.0092$ & $0.8168^{+0.0065}_{-0.0057}$ & $0.8163^{+0.0052}_{-0.0039}$\\
    $S_8$ & $0.835\pm 0.011$ & $0.8272\pm 0.0092$ & $0.8180\pm 0.0077$ & $0.8171\pm 0.0070$\\
    $H_0r_d~\mathrm{[km~s^{-1}]}$ & $9804^{+260}_{-160}$ & $10120\pm52$ & $10100\pm 49$ & $10100\pm 48$
    \enddata
\end{deluxetable}
\begin{figure}
    \centering
    \includegraphics[width=\linewidth]{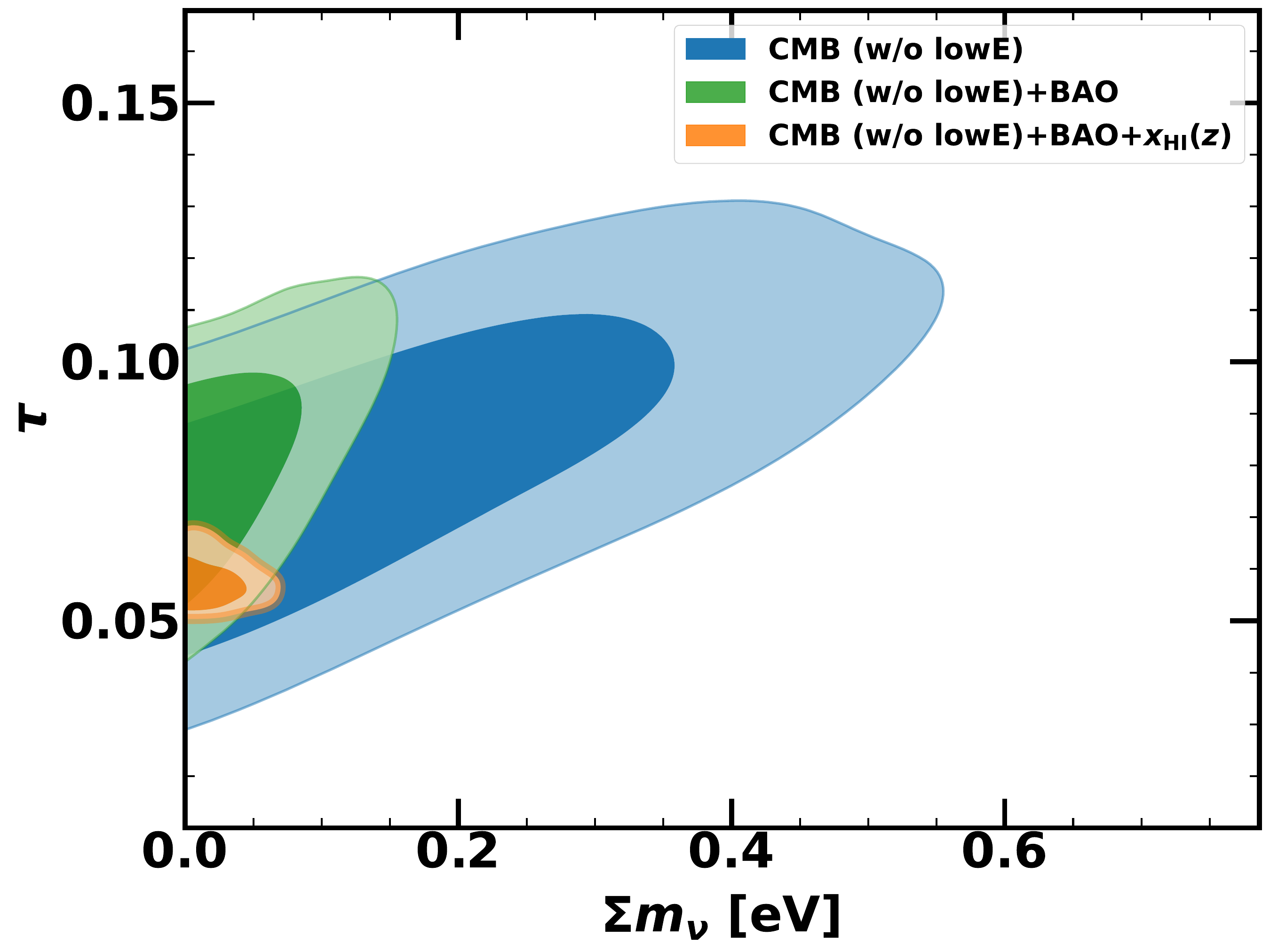}
    \caption{Marginalized posterior distributions of the ($\Sigma m_\nu$-$\tau$) parameters from CMB (w/o lowE) (blue), +BAO (green), and +$\xHI(z)$ (orange). A $95\%\,(99\%)$ upper limit of $\Sigma m_\nu<0.0550\,(0.0717)$ eV is obtained from CMB (w/o lowE)+BAO+$\xHI(z)$. A similar upper limit of $\Sigma m_\nu<0.0649\,(0.0906)$ eV (95\%\,(99\%)) is obtained from CMB (w/ lowE)+BAO. Contours show the $68\%$ and $95\%$ regions.}
    \label{fig:mnu_tau}
\end{figure}
\begin{figure}
    \centering
    \includegraphics[width=\linewidth]{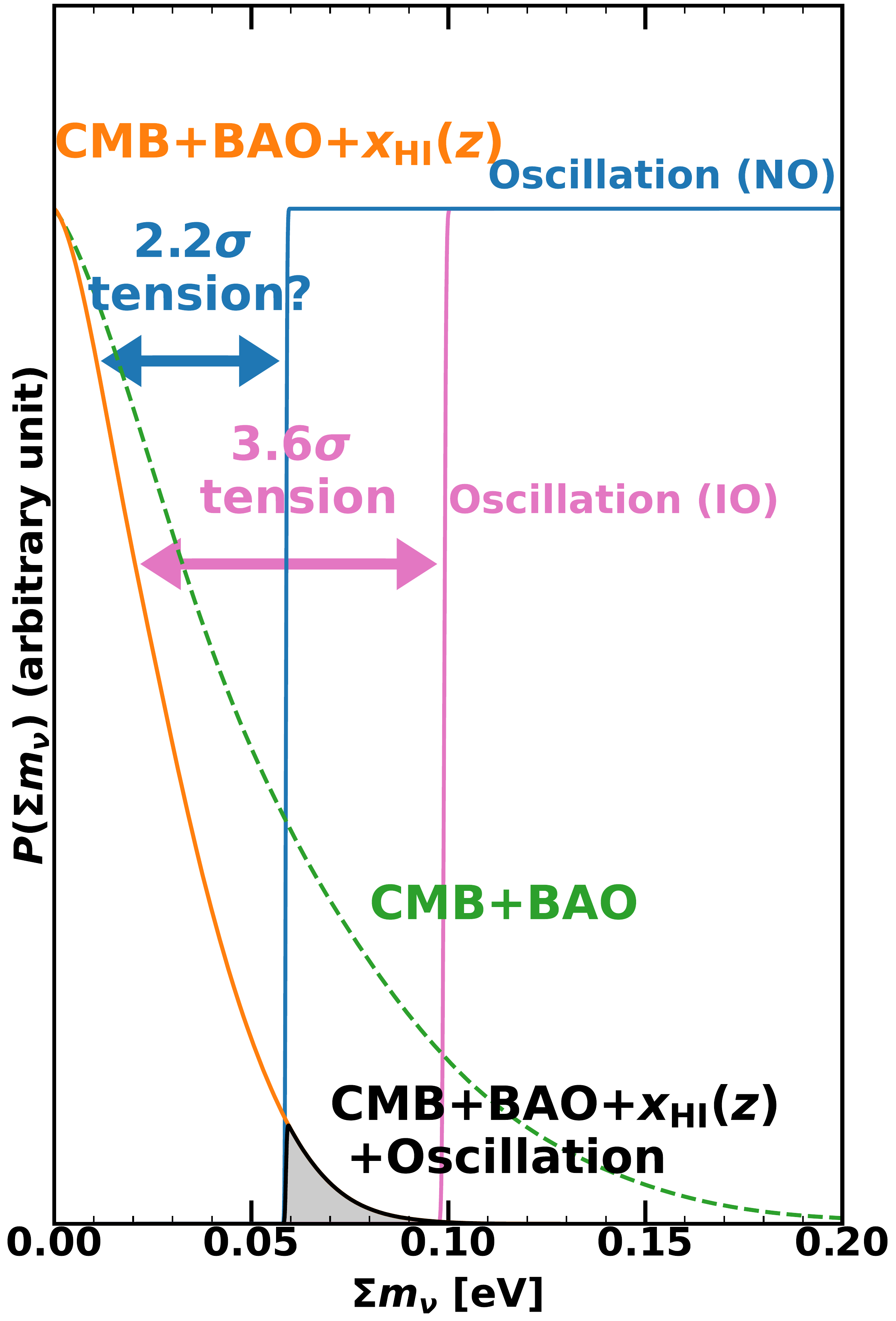}
    \caption{Marginalized posterior distributions of $\Sigma m_\nu$. An upper limit of $\Sigma m_\nu<0.0550\,(0.0717)$ eV (95\%\,(99\%) limit) is obtained from CMB (w/o lowE)+BAO+$\xHI(z)$ (orange). The blue (pink) line indicates the likelihood of $\Sigma m_\nu$ from neutrino oscillation experiments for normal (inverted) ordering, providing a lower limit on $\Sigma m_\nu$. The black line shows the probability distribution of $\Sigma m_\nu$ from the combination of cosmological data and neutrino oscillation constraints for normal ordering, which gives $\Sigma m_\nu=0.0594_{-0.0007}^{+0.0113}$ eV. The green dashed line indicates the posterior distribution from CMB (w/o lowE)+BAO. We note that our cosmological upper limit is in $2.2\sigma$ ($3.6\sigma$) tension with the oscillation-based lower limit for normal (inverted) ordering.}
    \label{fig:mnu_1d}
\end{figure}
\revise{Cosmological data sets including CMB and BAO measurements constrain the sum of neutrino masses $\Sigma m_\nu$, through their impact on the expansion history and growth of structure \citep{loverde24}.}
The sum of neutrino masses $\Sigma m_\nu$ is positively correlated with $\tau$ \revise{\citep{paoletti20, loverde24, elbers25_mnu, jhaveri25}}.
As shown in Section \ref{sec:cosmological_parameters}, raising $\tau$ increases the inferred $A_s$ and hence the amount of CMB lensing.
On the other hand, a larger $\Sigma m_\nu$ suppresses the growth of structure through free-streaming, thereby decreasing the CMB lensing signal \citep{du25}.
Therefore, it is important to investigate how the cosmic reionization history affects $\Sigma m_\nu$.
In Section \ref{sec:lcdm}, we fix the sum of neutrino masses to $\Sigma m_\nu=0.06$ eV.
In this section, we allow $\Sigma m_\nu$ to vary and derive an upper limit on $\Sigma m_\nu$ within $\Lambda$CDM cosmology.
We assume a flat prior on $\Sigma m_\nu$ over the range of $[0, 5]$ eV.
As in Section \ref{sec:dde}, we use three datasets to constrain $\Sigma m_\nu$: (i) CMB (w/o lowE), (ii) CMB (w/o lowE) and DESI DR2 BAO, and (iii) CMB (w/o lowE), DESI DR2 BAO, and reionization history $\xHI(z)$.
We follow the method in Section \ref{sec:method} and obtain constraints on cosmological parameters (Table \ref{tab:mnu}).
For comparison, we also show results from CMB (w/ lowE)+BAO based on the publicly available MCMC chains provided by the DESI Collaboration.
In Figure \ref{fig:mnu_tau}, we show the marginalized posterior distributions of the ($\Sigma m_\nu$-$\tau$) parameters.
The combination of CMB, BAO, and reionization history gives a tight upper limit on $\Sigma m_\nu$.
The derived $95\%$ (99\%) upper limit on $\Sigma m_\nu$ is
\begin{align}
    &\Sigma m_\nu<0.0550\,(0.0717)~\mathrm{eV}\notag\\
    &\qquad(95\%\,(99\%),\mathrm{CMB~(w/o~lowE)+BAO}+\xHI(z)).
\end{align}
This upper limit is similar to the lowE result, $\Sigma m_\nu<0.0649\,(0.0906)~\mathrm{eV}$ (95\% (99\%) limit, CMB (w/ lowE)+BAO).
Although cosmological data determine upper limits on $\Sigma m_\nu$, lower limits are obtained from neutrino oscillation experiments.
\citet{esteban24} report $95\%$ lower limits of $0.058~\mathrm{eV}<\Sigma m_\nu$ for normal ordering (NO) and $0.098~\mathrm{eV}<\Sigma m_\nu$ for inverted ordering (IO), based on multiple neutrino oscillation experiments with NuFit-6.0.
The cosmological upper limit is close to the lower limit for NO and well below that for IO.\par
First, we infer $\Sigma m_\nu$ by combining cosmological data with neutrino oscillation constraints.
In NO, the neutrino mass eigenvalues satisfy $m_1<m_2<m_3$.
From cosmological data, the sum of these eigenvalues,
\begin{align}
    \Sigma m_\nu=m_1+m_2+m_3,
\end{align}
is constrained.
On the other hand, neutrino oscillation experiments provide constraints on the squared mass differences,
\begin{align}
    \Delta m_{21}^2=m_2^2-m_1^2\quad\Delta m_{31}^2=m_3^2-m_1^2.
\end{align}
Here we treat $\Sigma m_\nu$, $\sqrt{\Delta m_{21}^2}$, and $\sqrt{\Delta m_{31}^2}$ as free parameters with the following prior distribution:
\begin{align}
    0~\mathrm{eV}<\Sigma m_\nu<5~\mathrm{eV}\notag\\
    0~\mathrm{eV}<\sqrt{\Delta m_{21}^2}<5~\mathrm{eV}\notag\\
    0~\mathrm{eV}<\sqrt{\Delta m_{31}^2}<5~\mathrm{eV}\notag\\
    \sqrt{\Delta m_{21}^2}<\sqrt{\Delta m_{31}^2}\notag\\
    \sqrt{\Delta m_{21}^2}+\sqrt{\Delta m_{31}^2}<\Sigma m_\nu\notag\\
    p(\Sigma m_\nu,\sqrt{\Delta m_{21}^2},\sqrt{\Delta m_{31}^2})\propto\Sigma m_\nu^{-2}.
\end{align}
This prior ensures that $m_1$ remains positive, and the marginalized prior $p(\Sigma m_\nu)$ is flat, making it consistent with the priors used in our cosmological analyses.
Using both cosmological and oscillation data ($d_c$ and $d_o$), the posterior distribution for $\Sigma m_\nu$ is given by
\begin{align}
    p&(\Sigma m_\nu|d_c,d_o)\notag\\
    &=\int\dd{\sqrt{\Delta m_{21}^2}}\dd{\sqrt{\Delta m_{31}^2}}\notag\\
    &\qquad\qquad p\qty(\Sigma m_\nu,\sqrt{\Delta m_{21}^2},\sqrt{\Delta m_{31}^2}|d_c,d_o)\notag\\
    &\propto\int\dd{\sqrt{\Delta m_{21}^2}}\dd{\sqrt{\Delta m_{31}^2}}p\qty(d_c|\Sigma m_\nu)\notag\\
    &\qquad\times p\qty(d_o|\sqrt{\Delta m_{21}^2},\sqrt{\Delta m_{31}^2})\notag\\
    &\qquad\times p\qty(\Sigma m_\nu,\sqrt{\Delta m_{21}^2},\sqrt{\Delta m_{31}^2}),
\end{align}
where $p\qty(d_c|\Sigma m_\nu)$ is the cosmological likelihood, $p\qty(d_o|\sqrt{\Delta m_{21}^2},\sqrt{\Delta m_{31}^2})$ is obtained from the $\chi^2$ values for neutrino mass differences (Section \ref{sec:bao}), and $p\qty(\Sigma m_\nu,\sqrt{\Delta m_{21}^2},\sqrt{\Delta m_{31}^2})$ is the prior distribution.
We evaluate the posterior distribution using the MCMC method with \texttt{emcee} (Figure \ref{fig:mnu_1d}).
The mode and $68\%$ credible interval of $\Sigma m_\nu$ for NO are
\begin{align}
    &\Sigma m_\nu=0.0594_{-0.0007}^{+0.0113}~\mathrm{eV}\notag\\&\quad(\mathrm{CMB~(w/o~lowE)+BAO}+\xHI(z)+\mathrm{oscillation}).
\end{align}
Although $\Sigma m_\nu$ can be inferred for IO in a similar manner, NO is preferred over IO with a Bayes factor of $110$.
The above constraint corresponds to a determination of $\Sigma m_\nu$ with $10\%$ precision.
The inferred value is close to the lower bound implied by neutrino oscillation experiments for NO, indicating that the neutrino mass spectrum is strongly hierarchical.
We obtain a lower bound on the ratio of the heaviest to the lightest masses, $m_3/m_1>3.7$ (95\% limit), implying that $m_3$ accounts for most of the total mass.
This result disfavors theoretical models that predict a quasi-degenerate spectrum, in which the three mass eigenvalues are nearly equal due to a small mass splitting arising from symmetry breaking \citep{lattanzi20}.\par
\begin{figure}
    \centering
    \includegraphics[width=\linewidth]{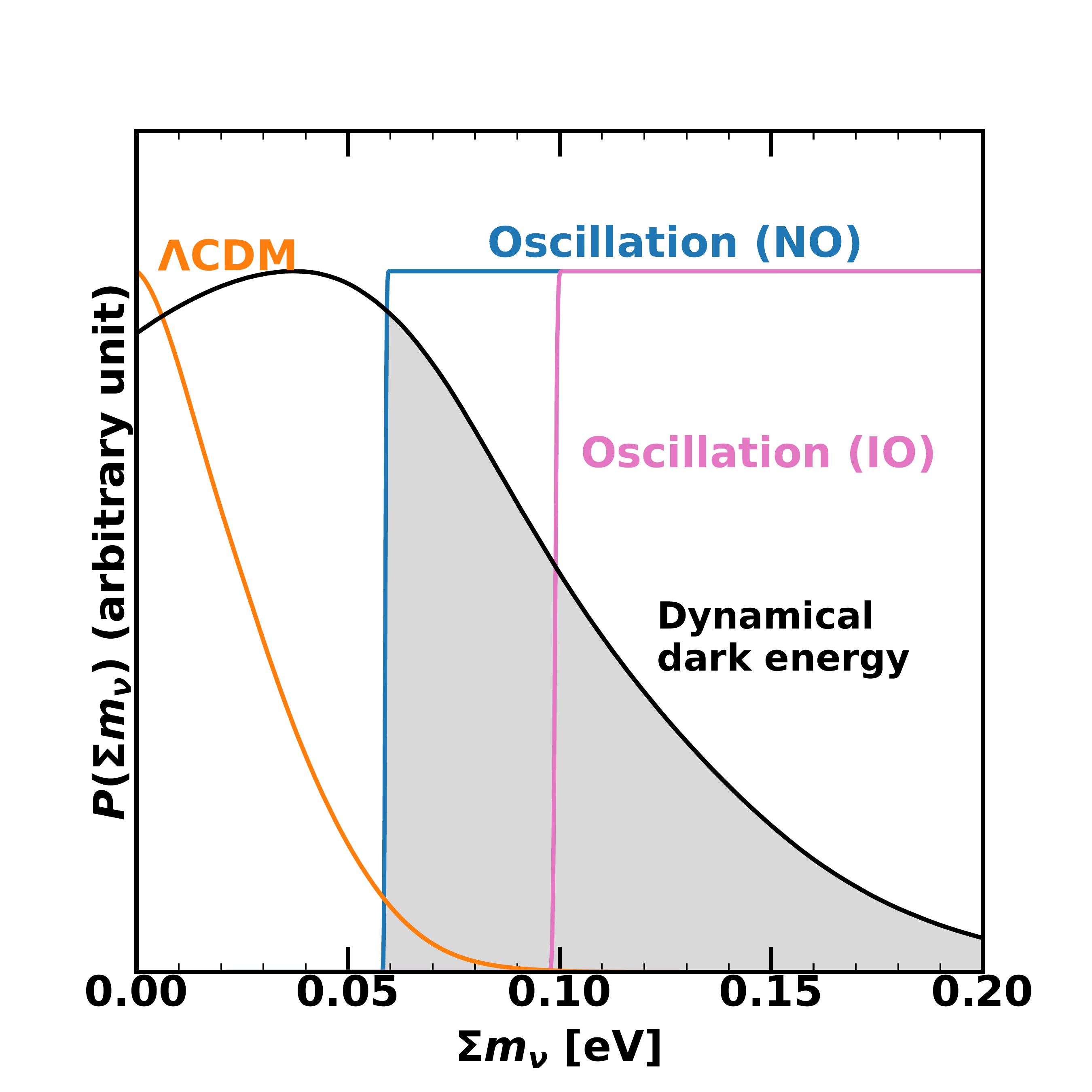}
    \caption{\revise{Marginalized posterior distributions of $\Sigma m_\nu$. An upper limit of $\Sigma m_\nu<0.141$~eV (95\% limit) is obtained from CMB (w/o lowE)+BAO+$\xHI(z)$ with the $w_0w_a$ model (black). The orange line indicates the result with the $\Lambda$CDM model. The blue (pink) line shows the likelihood of $\Sigma m_\nu$ from neutrino oscillation experiments for normal (inverted) ordering, providing a lower limit on $\Sigma m_\nu$. The $w_0w_a$ result is consistent with the neutrino oscillation constraints.}}
    \label{fig:mnu_1d_dde}
\end{figure}
However, we note that our cosmological constraint favors a very small $\Sigma m_\nu$ compared to the oscillation-based lower limit.
Even for NO, our $95\%$ upper limit of 0.0550~eV is smaller than the $95\%$ lower limit of 0.058~eV from neutrino oscillation experiments.
Our cosmological upper limit and the oscillation-based lower limit show a mild tension at the $2.2\sigma$ ($3.6\sigma$) level for NO (IO).
While such a neutrino mass tension has been noted in previous studies, it was argued that it could be resolved by removing the large-scale E-mode data \citep{jhaveri25}.
In this work, we identify the mild tension independently of large-scale E-mode measurements.
This suggests that the tension may originate from systematic errors in the BAO data or from new physics.
For example, neutrino decay may change the ``effective'' neutrino mass inferred from cosmological data \revise{and solve the neutrino mass tension \citep{abellan26}}.
Long-range forces between dark matter particles might even make the effective neutrino mass negative by enhancing matter clustering \citep{craig24}\revise{, although other work argues that similar interactions can instead suppress structure growth, implying a different impact on inferred $\Sigma m_\nu$ \citep{costa25}.}
In this study, we impose a prior that restricts the neutrino mass to positive values, and investigating the results obtained with a prior that allows negative masses is left for future work.
Alternatively, $\Lambda$CDM cosmology may require modification, and the dynamical dark energy model offers one possible solution \revise{\citep{ahlen25, elbers25_mnu}}.
\revise{When we treat $w_0$, $w_a$, and $\Sigma m_\nu$ as free parameters following Section \ref{sec:dde} and Section \ref{sec:mnu} and constrain these parameters using CMB (w/o lowE), BAO, and $\xHI(z)$ data, we obtain a $95\%$ upper limit of $\Sigma m_\nu<0.141~\mathrm{eV}$ (Figure \ref{fig:mnu_1d_dde}).
This upper limit is larger than the lower limit from neutrino oscillation experiments and is consistent with the CMB (w/ lowE) result of \citet{elbers25_mnu}, indicating that both the CMB-BAO tension and neutrino mass tension can be alleviated by dynamical dark energy.
Although the origin is still not clear, }
this neutrino mass tension provides an independent indication of possible physics beyond $\Lambda$CDM, in either particle physics or cosmology.\par
\revise{
We finally examine the impact of the possible absorption-model systematic in $\xHI(z)$, discussed in Section \ref{sec:sys}, on the neutrino
mass constraint.
Propagating this systematic uncertainty to the
cosmological upper limit, we obtain
\begin{equation}
\Sigma m_\nu < 0.055~{\rm eV}\ ({\rm stat.}) + 0.007~{\rm eV}\ ({\rm sys.})
\quad (95\%).
\end{equation}
Thus, even after accounting for this systematic uncertainty, the change in the upper limit is modest.
This is because the addition of the reionization-history information largely resolves the degeneracy between $\Sigma m_\nu$ and $\tau$, in the same way that it removes the $\Omega_m$--$\tau$ and $H_0$--$\tau$ degeneracies discussed in Section
\ref{sec:results_bao}.
The contours and one-dimensional posterior distributions in Figures \ref{fig:mnu_tau} and \ref{fig:mnu_1d} would therefore broaden only slightly if this systematic were
included, and the qualitative conclusions would remain unchanged.
In particular, the mild discrepancy between the cosmological upper limit and the oscillation-based lower limit for normal ordering remains at the $2.0\sigma$ level.
}

\section{Summary and Conclusion}\label{sec:summary}
In this paper, we present constraints on cosmological parameters using the redshift evolution of the neutral hydrogen fraction $\xHI(z)$.
To avoid potential biases in large-scale CMB E-mode polarization measurements due to instrumental noise or foreground subtraction, we use CMB power spectra that exclude large-scale E-mode measurements in our analyses.
In addition to the CMB power spectra, we use the latest measurements of $\xHI(z)$ based on Lyman-$\alpha$ forest data and QSO/galaxy Lyman-$\alpha$ damping-wing absorption measurements at $z\sim5-14$ to determine the CMB optical depth $\tau$.
These $\xHI(z)$ measurements are derived from ground-based optical and JWST observations.
Since many previous results are based on the same observational data, we avoid such duplication.
Our major findings are summarized below:
\begin{itemize}
    \item[1.] For the base-$\Lambda$CDM model, we find $\tau=0.0552^{+0.0019}_{-0.0026}$ from CMB (w/o lowE)+$\xHI(z)$. This result is consistent with the constraint from Planck CMB measurements including large-scale E-mode polarization.
    Without large-scale E-mode polarization, the addition of reionization history to CMB data improves the constraint on $\tau$ by a factor of $7$.
    \item[2.] There are also potential biases in $x_{\mathrm{HI}}(z)$ measurements due to astrophysical effects. We estimate the impact of a potential absorption-model systematic as $\tau=0.0552_{-0.0049}^{+0.0075}$ (systematics). Further investigations to quantify the impact of galaxy evolution are an important direction for future work.
    \item[3.] We resolve degeneracies in the $\tau$-$\Omega_m$ plane and find a $2.4\sigma$ tension with the DESI DR2 BAO results, which may suggest physics beyond $\Lambda$CDM, such as dynamical dark energy.
    \item[4.] Allowing the sum of neutrino masses $\Sigma m_\nu$ to vary, we derive an upper limit of $\Sigma m_\nu<0.0550\,(0.0717)$ eV ($95\%\,(99\%)$ limit) based on CMB (w/o lowE), DESI DR2 BAO, and reionization history $\xHI(z)$.
    This result strongly supports the normal mass ordering.
    Furthermore, combining these results with neutrino oscillation data yields $\Sigma m_\nu=0.0594_{-0.0007}^{+0.0113}$ eV.
    However, the cosmological upper limit and the oscillation-based lower limit show a $2.2\sigma$ mild tension, providing an independent indication of possible physics beyond $\Lambda$CDM.
\end{itemize}

\section*{Acknowledgments}
\revise{We thank the anonymous referee for the valuable comments that greatly improved this manuscript.}
We thank \revise{Richard Ellis,} Akio Inoue, Kazunori Kohri, Toshiya Namikawa, John Silverman, and Jun'ichi Yokoyama for valuable discussions on this work.
This work is based on observations obtained with Planck (\url{http://www.esa.int/Planck}), an ESA science mission with instruments and contributions directly funded by ESA Member States, NASA, and Canada.\par
This research used data obtained with the Dark Energy Spectroscopic Instrument (DESI). DESI construction and operations is managed by the Lawrence Berkeley National Laboratory. This material is based upon work supported by the U.S. Department of Energy, Office of Science, Office of High-Energy Physics, under Contract No. DE–AC02–05CH11231, and by the National Energy Research Scientific Computing Center, a DOE Office of Science User Facility under the same contract. Additional support for DESI was provided by the U.S. National Science Foundation (NSF), Division of Astronomical Sciences under Contract No. AST-0950945 to the NSF’s National Optical-Infrared Astronomy Research Laboratory; the Science and Technology Facilities Council of the United Kingdom; the Gordon and Betty Moore Foundation; the Heising-Simons Foundation; the French Alternative Energies and Atomic Energy Commission (CEA); the National Council of Humanities, Science and Technology of Mexico (CONAHCYT); the Ministry of Science and Innovation of Spain (MICINN), and by the DESI Member Institutions: \url{www.desi.lbl.gov/collaborating-institutions}. The DESI collaboration is honored to be permitted to conduct scientific research on I’oligam Du’ag (Kitt Peak), a mountain with particular significance to the Tohono O’odham Nation. Any opinions, findings, and conclusions or recommendations expressed in this material are those of the author(s) and do not necessarily reflect the views of the U.S. National Science Foundation, the U.S. Department of Energy, or any of the listed funding agencies.\par
This project used public archival data from the Dark Energy Survey (DES). Funding for the DES Projects has been provided by the U.S. Department of Energy, the U.S. National Science Foundation, the Ministry of Science and Education of Spain, the Science and Technology FacilitiesCouncil of the United Kingdom, the Higher Education Funding Council for England, the National Center for Supercomputing Applications at the University of Illinois at Urbana-Champaign, the Kavli Institute of Cosmological Physics at the University of Chicago, the Center for Cosmology and Astro-Particle Physics at the Ohio State University, the Mitchell Institute for Fundamental Physics and Astronomy at Texas A\&M University, Financiadora de Estudos e Projetos, Funda{\c c}{\~a}o Carlos Chagas Filho de Amparo {\`a} Pesquisa do Estado do Rio de Janeiro, Conselho Nacional de Desenvolvimento Cient{\'i}fico e Tecnol{\'o}gico and the Minist{\'e}rio da Ci{\^e}ncia, Tecnologia e Inova{\c c}{\~a}o, the Deutsche Forschungsgemeinschaft, and the Collaborating Institutions in the Dark Energy Survey. The Collaborating Institutions are Argonne National Laboratory, the University of California at Santa Cruz, the University of Cambridge, Centro de Investigaciones Energ{\'e}ticas, Medioambientales y Tecnol{\'o}gicas-Madrid, the University of Chicago, University College London, the DES-Brazil Consortium, the University of Edinburgh, the Eidgen{\"o}ssische Technische Hochschule (ETH) Z{\"u}rich,  Fermi National Accelerator Laboratory, the University of Illinois at Urbana-Champaign, the Institut de Ci{\`e}ncies de l'Espai (IEEC/CSIC), the Institut de F{\'i}sica d'Altes Energies, Lawrence Berkeley National Laboratory, the Ludwig-Maximilians Universit{\"a}t M{\"u}nchen and the associated Excellence Cluster Universe, the University of Michigan, the National Optical Astronomy Observatory, the University of Nottingham, The Ohio State University, the OzDES Membership Consortium, the University of Pennsylvania, the University of Portsmouth, SLAC National Accelerator Laboratory, Stanford University, the University of Sussex, and Texas A\&M University. Based in part on observations at Cerro Tololo Inter-American Observatory, National Optical Astronomy Observatory, which is operated by the Association of Universities for Research in Astronomy (AURA) under a cooperative agreement with the National Science Foundation.\par
Based on data obtained from the ESO Science Archive Facility with DOI: \url{https://doi.org/10.18727/archive/37}, and \url{https://doi.eso.org/10.18727/archive/59} and on data products produced by the KiDS consortium. The KiDS production team acknowledges support from: Deutsche Forschungsgemeinschaft, ERC, NOVA and NWO-M grants; Target; the University of Padova, and the University Federico II (Naples).\par
The Hyper Suprime-Cam (HSC) collaboration includes the astronomical communities of Japan and Taiwan, and Princeton University. The HSC instrumentation and software were developed by the National Astronomical Observatory of Japan (NAOJ), the Kavli Institute for the Physics and Mathematics of the Universe (Kavli IPMU), the University of Tokyo, the High Energy Accelerator Research Organization (KEK), the Academia Sinica Institute for Astronomy and Astrophysics in Taiwan (ASIAA), and Princeton University. Funding was contributed by the FIRST program from Japanese Cabinet Office, the Ministry of Education, Culture, Sports, Science and Technology (MEXT), the Japan Society for the Promotion of Science (JSPS), Japan Science and Technology Agency (JST), the Toray Science Foundation, NAOJ, Kavli IPMU, KEK, ASIAA, and Princeton University. 
This paper makes use of software developed for the Large Synoptic Survey Telescope. We thank the LSST Project for making their code available as free software at \url{http://dm.lsst.org}.
The Pan-STARRS1 Surveys (PS1) have been made possible through contributions of the Institute for Astronomy, the University of Hawaii, the Pan-STARRS Project Office, the Max-Planck Society and its participating institutes, the Max Planck Institute for Astronomy, Heidelberg and the Max Planck Institute for Extraterrestrial Physics, Garching, The Johns Hopkins University, Durham University, the University of Edinburgh, Queen’s University Belfast, the Harvard-Smithsonian Center for Astrophysics, the Las Cumbres Observatory Global Telescope Network Incorporated, the National Central University of Taiwan, the Space Telescope Science Institute, the National Aeronautics and Space Administration under Grant No. NNX08AR22G issued through the Planetary Science Division of the NASA Science Mission Directorate, the National Science Foundation under Grant No. AST-1238877, the University of Maryland, and Eotvos Lorand University (ELTE) and the Los Alamos National Laboratory.
Based (in part) on data collected at the Subaru Telescope and retrieved from the HSC data archive system, which is operated by Subaru Telescope and Astronomy Data Center at National Astronomical Observatory of Japan.\par
This publication is based on work supported by the World Premier International Research Center Initiative (WPI Initiative), MEXT, Japan, KAKENHI (25H00674) through the Japan Society for the Promotion of Science.
This work is supported by the joint research program of the Institute for Cosmic Ray Research (ICRR), the University of Tokyo.
YK and FN are supported by Forefront Physics and Mathematics Program to Drive Transformation (FoPM), a World-leading Innovative Graduate Study (WINGS) Program, the University of Tokyo.
We acknowledge support from KAKENHI Grant Nos. 24KJ0668 (FN), 24KJ0575 (AM), and 25KJ0828 (MN) through Japan Society for the Promotion of Science (JSPS). FN also thank support from the ANRI Fellowship.
\revise{The authors acknowledge the use of ChatGPT (OpenAI, GPT-5.5) and Codex (OpenAI, Codex 5.5) to assist with language editing and code development.
All AI-assisted outputs were carefully reviewed and validated by the authors.
The authors take full responsibility for all analyses, interpretations, and conclusions presented in this work.}
\software{CAMB \citep{lewis00, howlett12}, Cobaya \citep{torrado19, torrado21}, emcee \citep{foremanmackey13}, GetDist \citep{lewis25}, Matplotlib \citep{hunter07}, NumPy \citep{harris20}, SciPy \citep{virtanen20}}

\appendix
\section{Details of Gaussian Process}\label{sec:gp}
We use constraints on the $\xHI$ values at redshifts $\vb{z}=(z_1, z_2, ..., z_n)$ to reconstruct the reionization history (Section \ref{sec:crh}).
The probability distribution of $\xHI$ at $z=z_i$ is denoted as $p_i(\xHI)$, and the full set of constraints on $\xHI$ is expressed as $\{\xHI\}$.
Although Gaussian Process regression describes functions whose outputs can take any real value, the neutral fraction $\xHI$ is physically constrained to the range of $0\le\xHI\le1$.
To enforce this constraint, we introduce a latent function $f(z)\in\mathbb{R}$ and perform Gaussian process regression on $f(z)$ instead of $\xHI(z)$.
The conversion from $f(z)$ to $\xHI(z)$ is achieved using a sigmoid function,
\begin{align}\label{equ:sigmoid}
    \xHI=\frac{1}{1+\exp(-f)}.
\end{align}
\par
We assume a Gaussian process prior (i.e., a multivariate normal distribution) for the random variables $\vb{f}\equiv(f(z_1), f(z_2), ..., f(z_n))$ at redshifts $z=z_1, z_2, ..., z_n$:
\begin{align}
    f(\vb{z})\sim\mathcal{N}(\mu(\vb{z}), k(\vb{z}, \vb{z}')).
\end{align}
For the mean function of the Gaussian process prior, we adopt $\mu(z)=z-7.67$ to reflect the monotonically increasing nature of $\xHI$.
The value $7.67$ corresponds to the midpoint of reionization derived from the Planck TTTEEE+lowE+lensing results \citep{planck18_6}.
For the covariance of the Gaussian process prior, we use the radial basis function (RBF) kernel,
\begin{align}
    k(z,z')=s^2&\exp\qty(-\frac{(\log_{10}(1+z)-\log_{10}(1+z'))^2}{2r^2})\notag\\
    &+j\delta(z,z').
\end{align}
Here, $j=10^{-6}$ is included for numerical stability, and $s$ and $r$ are hyperparameters of the RBF kernel.
We determine these hyperparameters by maximizing the marginalized likelihood,
\begin{align}
    L(s,r|\{\xHI\})=\int\dd{\vb{f}}L(s,r,\vb{f}|\{\xHI\}).
\end{align}
We obtain $s=0.83$ and $r=0.076$ with this procedure.\par
Using these hyperparameters, we perform Gaussian process regression.
The posterior distribution is expressed as
\begin{align}
    \log&{p(\vb{f}|\{\xHI\})}\notag\\
    &=-\frac{1}{2}(\vb{f}-\mu(\vb{z}))^TK(\vb{z},\vb{z}')^{-1}(\vb{f}-\mu(\vb{z}))\notag\\
    &\quad+\sum_{i=1}^n\log(p_i(\xHI(f(z_i)))\dv{\xHI}{f}\qty(f(z_i)))+const.
\end{align}
We evaluate the posterior distribution of $\vb{f}=(f(z_1), ..., f(z_n))$ using the MCMC method with \texttt{emcee} \citep{foremanmackey13}.
Given the posterior distribution at $z=z_1,...,z_n$, we compute the values $\vb{f}^\star$ at new redshift points $\vb{z}^\star=(z_1^\star,...,z_{n'}^\star)$ using the Gaussian process prior:
\begin{align}
    p(\vb{f}^\star)|\{\xHI\})=\int\dd{\vb{f}}p(\vb{f}^\star|\vb{f})p(\vb{f}|\{\xHI\}),
\end{align}
where the conditional probability $p(\vb{f}^\star|\vb{f})$ is given by:
\begin{align}
    \vb{f}^\star|\vb{f}\sim\mathcal{N}&(\mu(\vb{z}^\star)+k(\vb{z}^\star,\vb{z})k(\vb{z},\vb{z})^{-1}(\vb{f}-\mu(\vb{z})),\notag\\
    &k(\vb{z}^\star,\vb{z}^\star)-k(\vb{z}^\star,\vb{z})k(\vb{z},\vb{z})^{-1}k(\vb{z},\vb{z}^\star)).
\end{align}
By converting $\vb{f}^\star$ to $\xHI$ using Equation (\ref{equ:sigmoid}), we obtain the posterior distribution of $\xHI$ at new redshift points, $p(\xHI(\vb{z}^\star)|\{\xHI\})$.

\section{Optical Depth Measurement from Other Datasets of the Reionization History}\label{sec:TM}
\begin{figure}
    \centering
    \includegraphics[width=\linewidth]{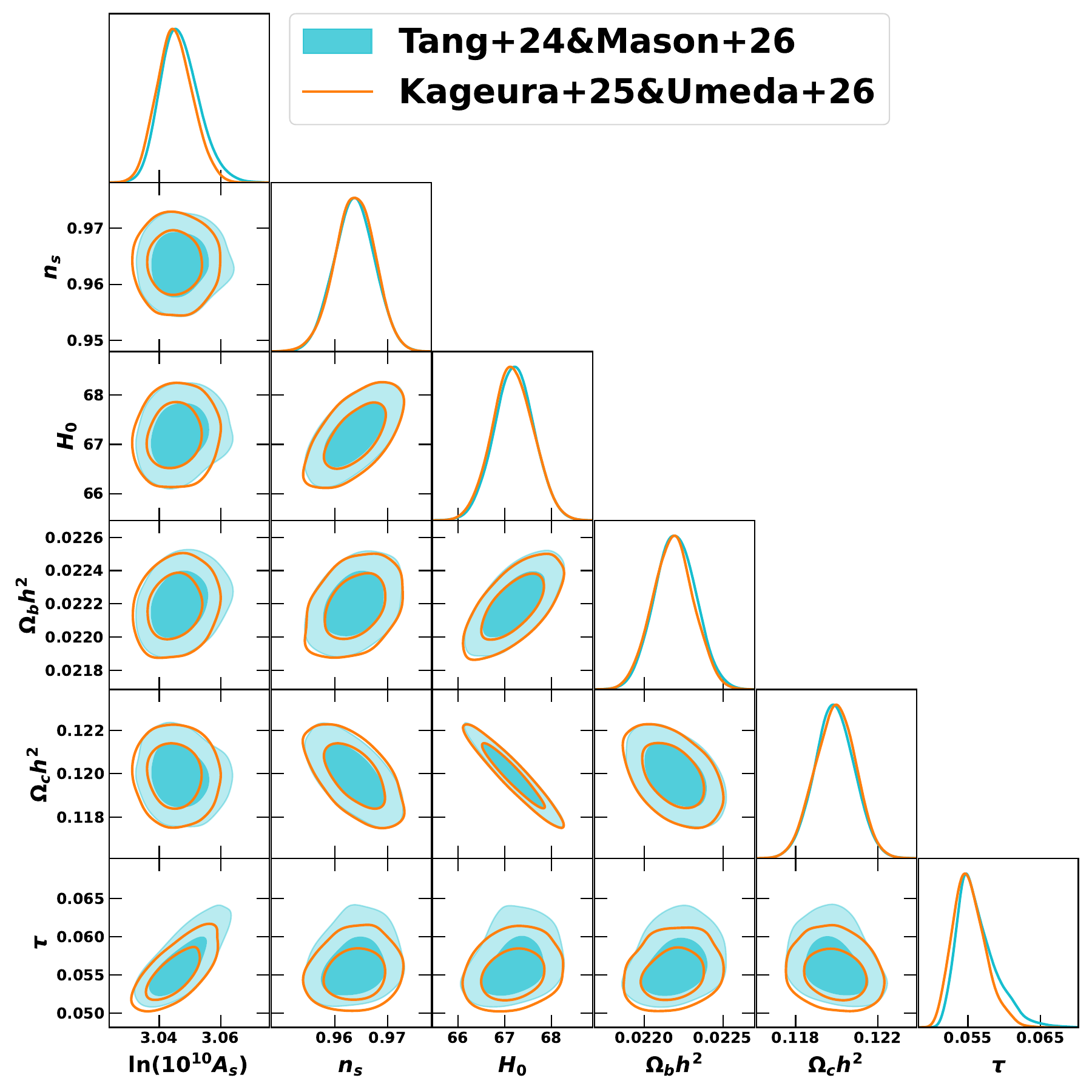}
    \caption{\revise{Constraints on cosmological parameters and $\tau$ for the $\Lambda$CDM model. The orange open contours represent constraints derived from CMB (w/o lowE)+$\xHI(z)$ with our fiducial reionization datasets presented in Section \ref{sec:crh}, while the cyan filled contours show the results using JWST $\xHI$ measurements by \citet{tang24} and \citet{mason25} instead of \citet{kageura25} and \citet{umeda25b}. The two sets of contours are consistent with each other, suggesting that changes in the reionization history datasets do not significantly impact the estimates of cosmological parameters.}}
    \label{fig:TM}
\end{figure}
\revise{
There are several JWST studies using Ly$\alpha$ damping-wing absorption of Ly$\alpha$ emission and rest-UV continuum.
Since these studies are partially based on the same observational data, they are not statistically independent.
We choose \citet{umeda25b} and \citet{kageura25} for JWST Ly$\alpha$ damping-wing absorption of LBGs and the evolution of the Ly$\alpha$ EW distribution, respectively.
In this section, we evaluate how the result on $\tau$ would change with different datasets.
We use the JWST $\xHI$ measurements by \citet{tang24} and \citet{mason25} instead of \citet{kageura25} and \citet{umeda25b} to derive $\tau$ and cosmological parameters in the $\Lambda$CDM framework.
We show the posterior distributions of the cosmological parameters and $\tau$ in Figure \ref{fig:TM}.
The derived value of the optical depth is $\tau=0.0563_{-0.0032}^{+0.0017}$, which is consistent with the result presented in Section \ref{sec:cosmological_parameters}.
Therefore we conclude that changes on the reionization history datasets do not significantly impact the estimate of cosmological parameters.}

\section{Constraints on $\Omega_\Lambda$}\label{sec:omegal}
\begin{figure}
    \centering
    \includegraphics[width=\linewidth]{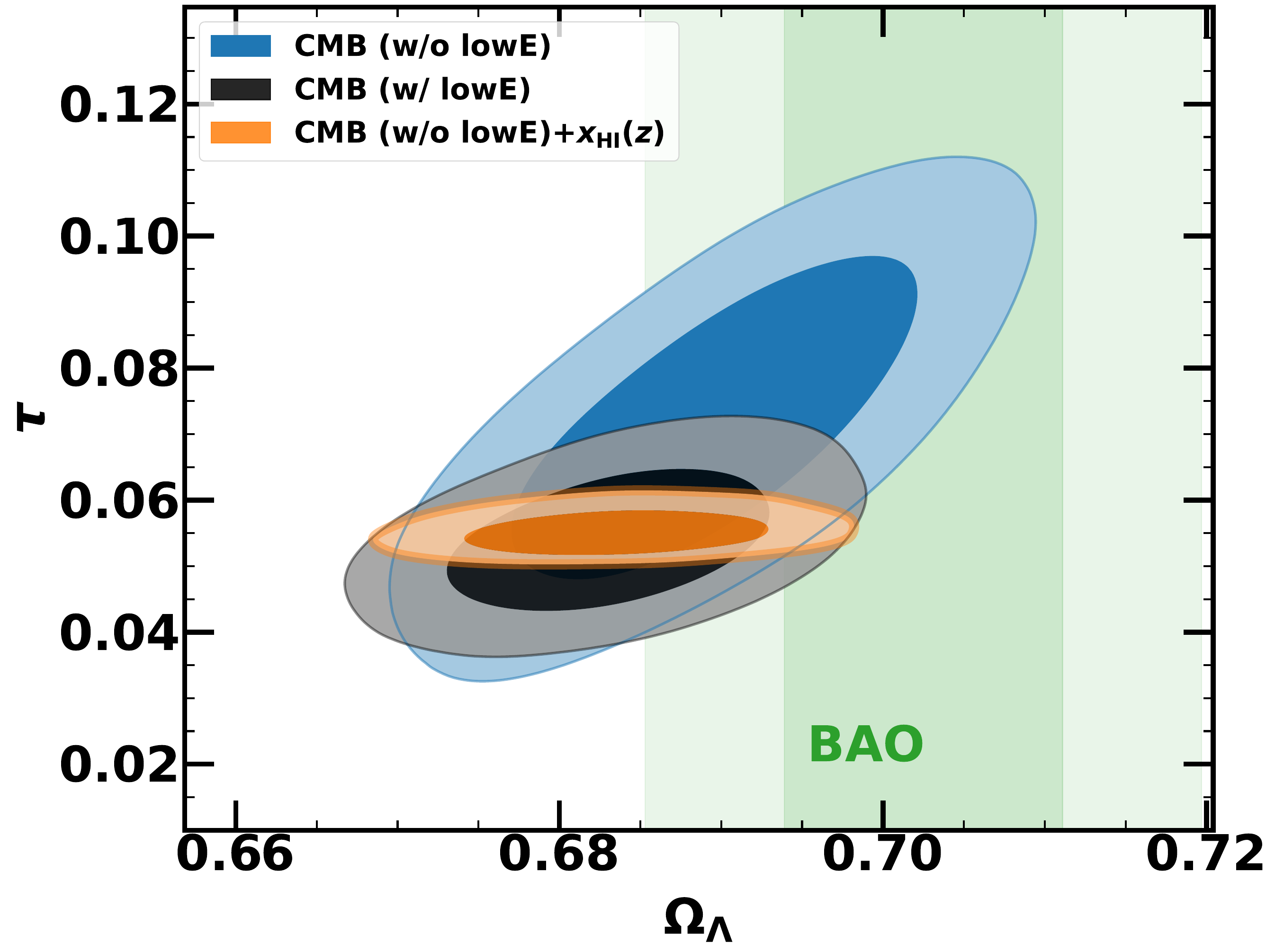}
    \caption{\revise{Constraints on $\Omega_\Lambda$ for the $\Lambda$CDM model. The blue, black, and orange contours show results based on CMB (w/o lowE), CMB (w/ lowE), and CMB (w/o lowE)+$\xHI(z)$, respectively. The green region indicates constraints from DESI DR2 BAO measurements. Contours show the 68\% and 95\% regions.}}
    \label{fig:omegal_tau}
\end{figure}

\revise{For completeness, Figure \ref{fig:omegal_tau} shows the constraints on $\Omega_\Lambda$. In flat $\Lambda$CDM, this is
equivalent to $\Omega_m$ because $\Omega_\Lambda=1-\Omega_m$.
}

\section{Cosmological Parameters Based on CMB (w/ lowE)+$\xHI(z)$}\label{sec:lowE_xHI}
\begin{figure}
    \centering
    \includegraphics[width=\linewidth]{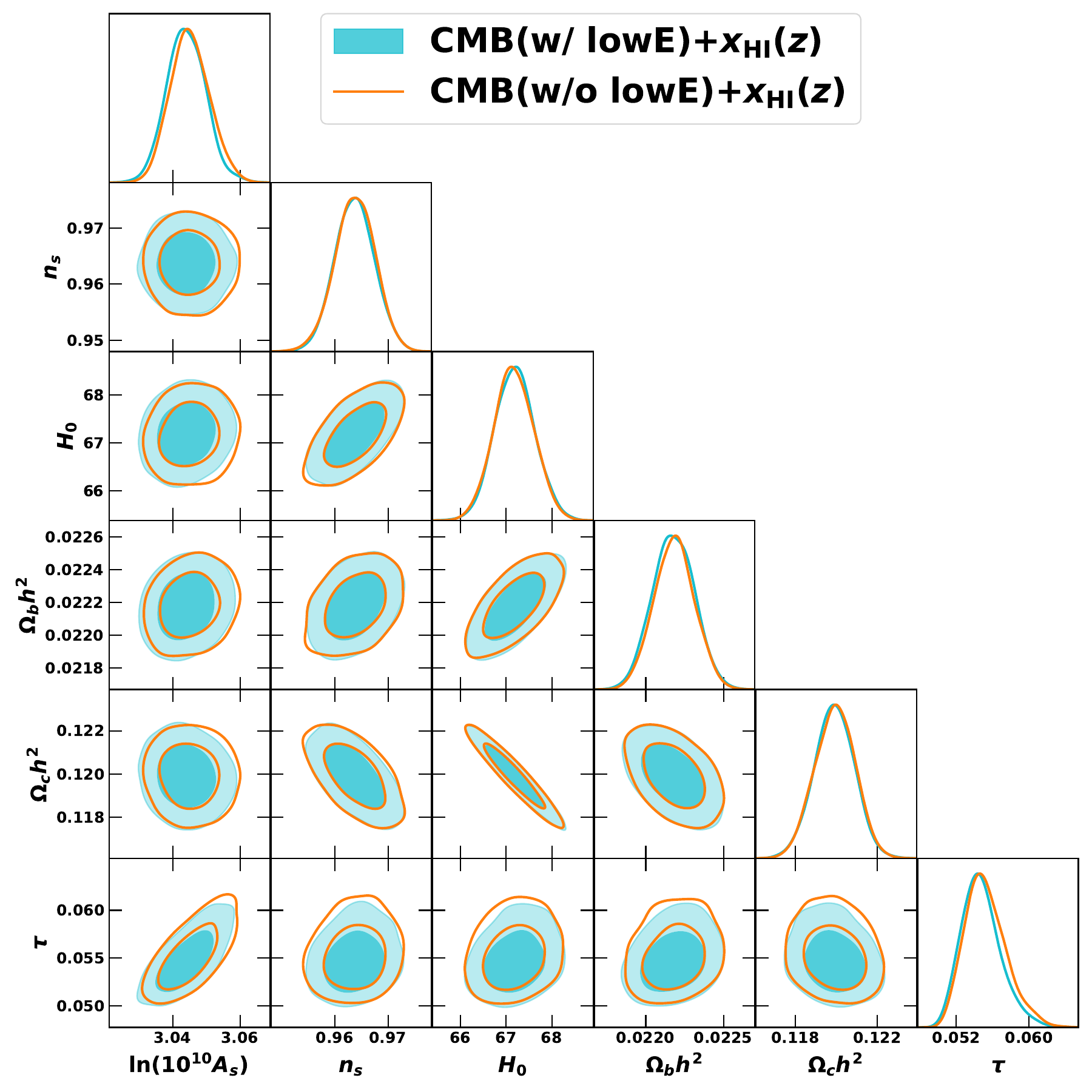}
    \caption{\revise{Constraints on cosmological parameters and $\tau$ for the $\Lambda$CDM model. The orange open (cyan filled) contours represent constraints derived from CMB (w/o lowE)+$\xHI(z)$ (CMB (w/ lowE)+$\xHI(z)$). The dark and light contours show the 68\% and 95\% regions, respectively. Since the lowE and reionization-history constraints are consistent with each other, the two sets of contours are nearly identical.}}
    \label{fig:contour_full}
\end{figure}
\revise{In Section \ref{sec:lcdm}, we constrain cosmological parameters based on CMB (w/o lowE) + $\xHI(z)$ to avoid possible systematics in lowE.
For comparison, we also constrain $\Lambda$CDM cosmological parameters using both lowE and the reionization history (i.e., CMB (w/ lowE)+$\xHI(z)$).
In addition to CMB likelihoods presented in Table \ref{tab:cmb}, we use the \texttt{SimAll} likelihood from the Planck PR3 release for the low-$l$ EE power spectrum \citep{planck18_5}.
In Figure \ref{fig:contour_full}, we show the posterior distributions of the cosmological parameters and $\tau$.
The optical depth inferred from CMB (w/ lowE)+$\xHI(z)$ is $\tau=0.0548_{-0.0025}^{+0.0017}$, with both the mean and $68\%$ credible interval nearly identical to those obtained without lowE.
Since our $\tau$ value based on the reionization history is consistent with the Planck lowE measurement, the inclusion of lowE does not significantly alter the result of $\tau$ and other cosmological parameters.}

\bibliography{kageura26}{}
\bibliographystyle{aasjournal}



\end{document}